\documentclass[prl,twocolumn,showpacs,showkeys,floatfix]{revtex4}
\usepackage{graphicx}
\usepackage{subfigure}
\usepackage{xcolor}
\usepackage{multirow}
\usepackage{setspace}
\usepackage{dcolumn}
\usepackage{amsmath}
\usepackage{float}
\usepackage{tablefootnote}
\usepackage{longtable}
\usepackage{amssymb}
\definecolor{calpolypomonagreen}{rgb}{0.12, 0.3, 0.17}
\definecolor{blue}{rgb}{0.0, 0.0, 1.0}
\definecolor{bleudefrance}{rgb}{0.19, 0.55, 0.91}
\definecolor{red}{rgb}{1.0, 0.0, 0.0}
\definecolor{byzantine}{rgb}{0.74, 0.2, 0.64}
\definecolor{cadmiumgreen}{rgb}{0.0, 0.42, 0.24}
\definecolor{yellow}{rgb}{0.99, 0.93, 0.0}

\begin{document}
\title{The Origin of the Giant Dipole Resonance}

\author{R.B.~Firestone}
\affiliation{University of California, Department of Nuclear Engineering,  Berkeley, CA 94720, USA}

\date{\today}

\begin{abstract}

The Giant Dipole Resonance (GDR) is shown to result from an increase in level density at a harmonic oscillator gap excitation energy of $2\hbar\omega$.  The photonuclear reaction photon strength populating the GDR follows a Lorentzian distribution with three parameters, average energy, $\overline{E}$, total cross section, $\sigma_T$, and average width, $\overline{\Gamma}$.  A fit to 148 experimental measurements of the GDR photon strength with A=2-239 gives ${\overline{E}=2\hbar\omega=2(47.34\pm0.27)(A^{-1/3}-A^{-2/3})}$, in close agreement with the harmonic oscillator energies determined from nuclear radii.  The total GDR cross section follows a mass dependent increase in level density and is given by $\sigma_T=(0.483\pm0.005)A^{4/3}$.  The average GDR width follows the energy dependence $\overline{\Gamma}=(1.110\pm0.015)E^{1/2}$ implied by the Lorentzian distribution.  For deformed nuclei the GDR separates into two peaks, each with its own GDR parameters.  The energy separation of the two peaks is proportional to the ground state deformation, $|\beta_2|$, and given by $E_2-E_1=(11.1\pm0.3)|\beta_2|$.  The cross sections and widths of the two peaks follow the asymmetric relation $\sigma_2/\sigma_1=\Gamma_2/\Gamma_1=1.50\pm0.07$.  This asymmetry is consistent with Nilsson model calculations where the two peaks correspond to transitions to states with oblate and prolate deformations respectively.  The association of the GDR with level density requires that additional resonances occur at all higher oscillator gaps.  It is shown that the Giant Quadrupole (GQR), Isoscalar GDR (ISGDR), Giant Monopole (GMR), and Giant Octupole (GOR) resonances all occur at excitations of ${E=2-4\hbar\omega}$.  These results rule out the common assumption that the GDR is a collective enhancement of photon strength due to the counter motion of protons and neutrons.

\end{abstract}


\maketitle

\section{Introduction}

In 1948 Goldhaber and Teller~\cite{Goldhaber48} proposed that the Giant Dipole Resonance (GDR) is due to "$\gamma$-rays exciting a motion in the nucleus in which the bulk of the protons move in one direction while the neutrons move in the opposite direction". Although this picture of nucleons collectively sloshing about the nucleus is simplistic, it remains the predominant explanation for a large, high energy peak observed in the cross sections of photonuclear reactions.  Photonuclear experiments measure the photoexcitation cross sections $\sigma(\gamma,n)$ populating states above the neutron separation energy, $S_n$.  The $\gamma$-rays have predominantly E1 multipolarity and the photon strength , $f(E_{\gamma})$, is related to the cross section by detailed balance as was shown by Uhl and Kopecky~\cite{Uhl94} and given in Eq.~\ref{EQ1} where $\rho(E_x)$ is the average level density,
\begin{equation}
\label{EQ1}
f(E_{\gamma}) = \frac{\sigma_{\gamma}(E_x)} {3\pi^2 \hbar^2 c^2 E_{\gamma}} \propto \rho(E_x)\frac{\Gamma(E_{\gamma})}{E_{\gamma}^3}
\end{equation}
$\Gamma(E_{\gamma})/E_{\gamma}^3$ is the average reduced level width, and the level excitation energy $E_x=E_{\gamma}$.  The contribution of M1 and E2 multipolarities is $\lesssim$1\%~\cite{Kopecky90}. The photon strength is the product of level density and reduced level width so the peak in the photon strength distribution can be due to fluctuations in either quantity.

\begin{figure}[!ht]
  \centering
    \includegraphics[width=8cm]{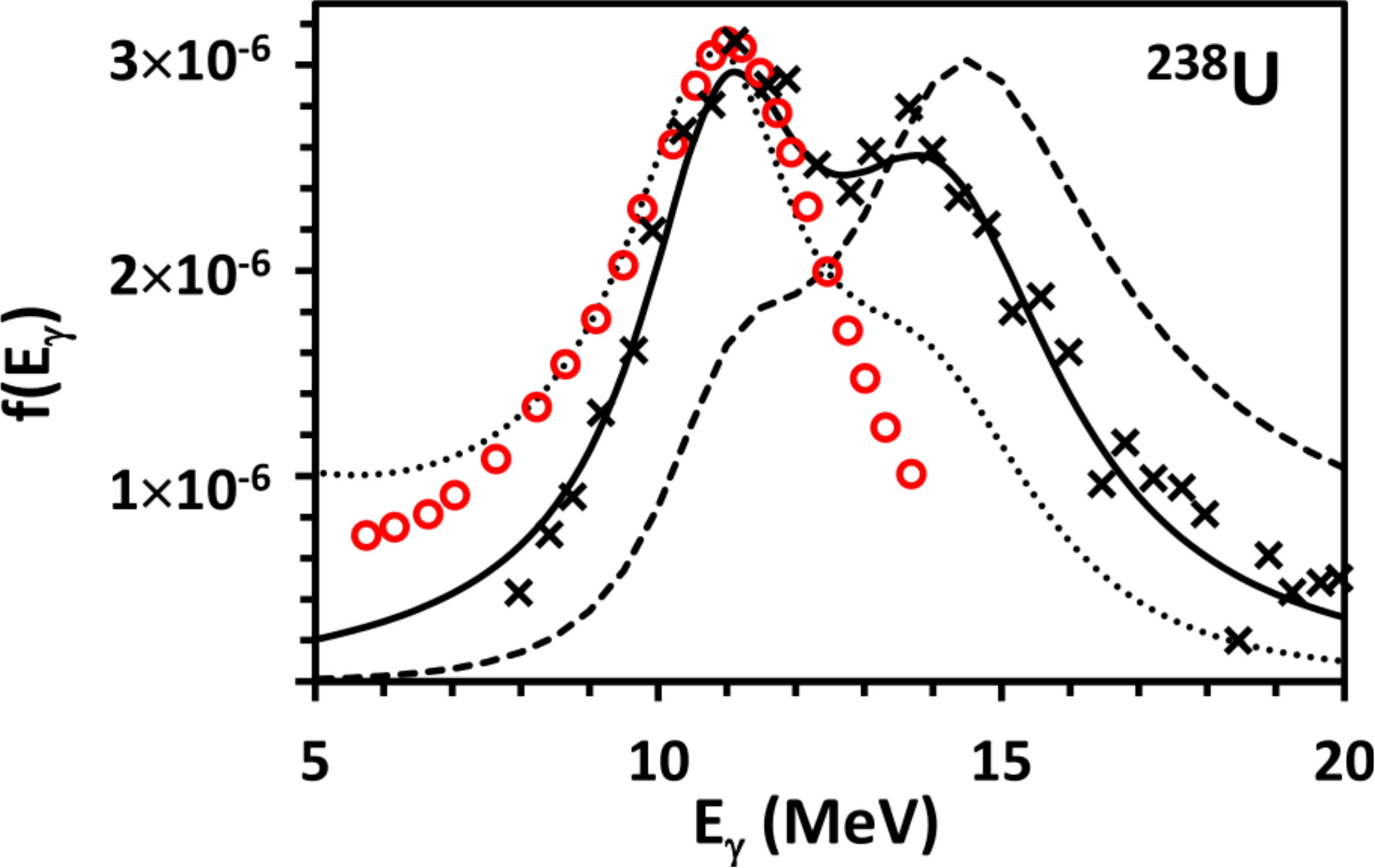}
  \caption{Comparison of $f_{\gamma}(^{238}U)$ photon strength (\textbf{---}) with experiment (\textbf{x}), $f_{\gamma}(^{238}U) \times E_{\gamma}^3 /2400$ (\textbf{-~-~-}) removing energy dependence, $f_{\gamma}(^{238}U) \times E_{\gamma}^{-2} \times 125$ (\textbf{$\cdot\cdot\cdot$}) for E2 transitions, and experimental GQR data$\times 60$~\cite{Arruda80}, shifted by 0.75 MeV.}
  \label{GDRHO}
\end{figure}

The shape of the GDR photon strength is well described by a simple Lorentzian distribution.  It was parameterized by Brink and Axel (BA)~\cite{Brink55,Axel62} as shown in Eq.~\ref{EQ2} where $E_i$ is the GDR energy in MeV,  $\Gamma_i$ is its width in MeV, and $\sigma_i$ is the photonuclear cross section in mb.
\begin{equation}
f(E_{\gamma}) = \frac{1}{3(\pi \hbar c)^2}\sum_{i=1}^{i=2} \frac{\sigma_{i} E_{\gamma} \Gamma_{i}^2} {(E_{\gamma}^2-E_{i}^2)^2+E_{\gamma}^2\Gamma_{i}^2}
\label{EQ2}
\end{equation}
The summation is due to the splitting of GDR into two peaks in deformed nuclei leading to two sets of GDR peak parameters.  Recommended experimental GDR parameters for numerous nuclei have been published by the International Atomic Energy Agency (IAEA)~\cite{Varlamov99,Kawano19}.  An example of the photon strength calculated for $^{238}$U and compared with experiment~\cite{Varlamov99} is shown in Fig~\ref{GDRHO}.

The BA formulation describes the photon strength for a single nuclear reaction, e.g. ($\gamma$,n) or ($\gamma$,f).  If multiple reactions are energetically possible then $f(E_{\gamma})\rightarrow\sum_{i} f_i(E_{\gamma})$ where $f_i(E_{\gamma})$ is the contribution to the photon strength for each reaction.  At low energies only the $(\gamma,\gamma')$ reaction is possible and the BA formulation is not applicable.  Above the neutron separation energy, $S_n$, the $(\gamma,n)$ reaction becomes available and $f(E_{\gamma})=f_{\gamma,\gamma'}(E_{\gamma})+f_{\gamma,n}(E_{\gamma})$ although the BA formulation gives only the $(\gamma,n)$ component of the photon strength.  Well above $S_n$ $f_{\gamma,n}(E_{\gamma})\gg\Gamma_{\gamma,\gamma'}(E_{\gamma})$ so the BA formulation becomes asymptotically valid.  Nevertheless, the contribution of additional reactions, e.g. $(\gamma,2n)$, at higher energies will invalidate the BA formulation.

\section{Brink-Axel GDR photon strength parameters}

Experimental Brink-Axel GDR photon strength parameters have been compiled in the Reference Input Parameter Library (RIPL-3)~\cite{Capote09}.  They were generated from fits to experimental photonuclear data although the values from different experiments on the same target are often discrepant.  Recently a recommended set of BA photon strength parameters has been proposed~\cite{Kawano19}.  The systematic trends of these parameters can be expressed as simple functions of mass, $A$, and nuclear deformation, $\beta_2$ as is discussed below.

\subsection{GDR energy}

\begin{figure}[t]
\centering
    \includegraphics[width=8cm]{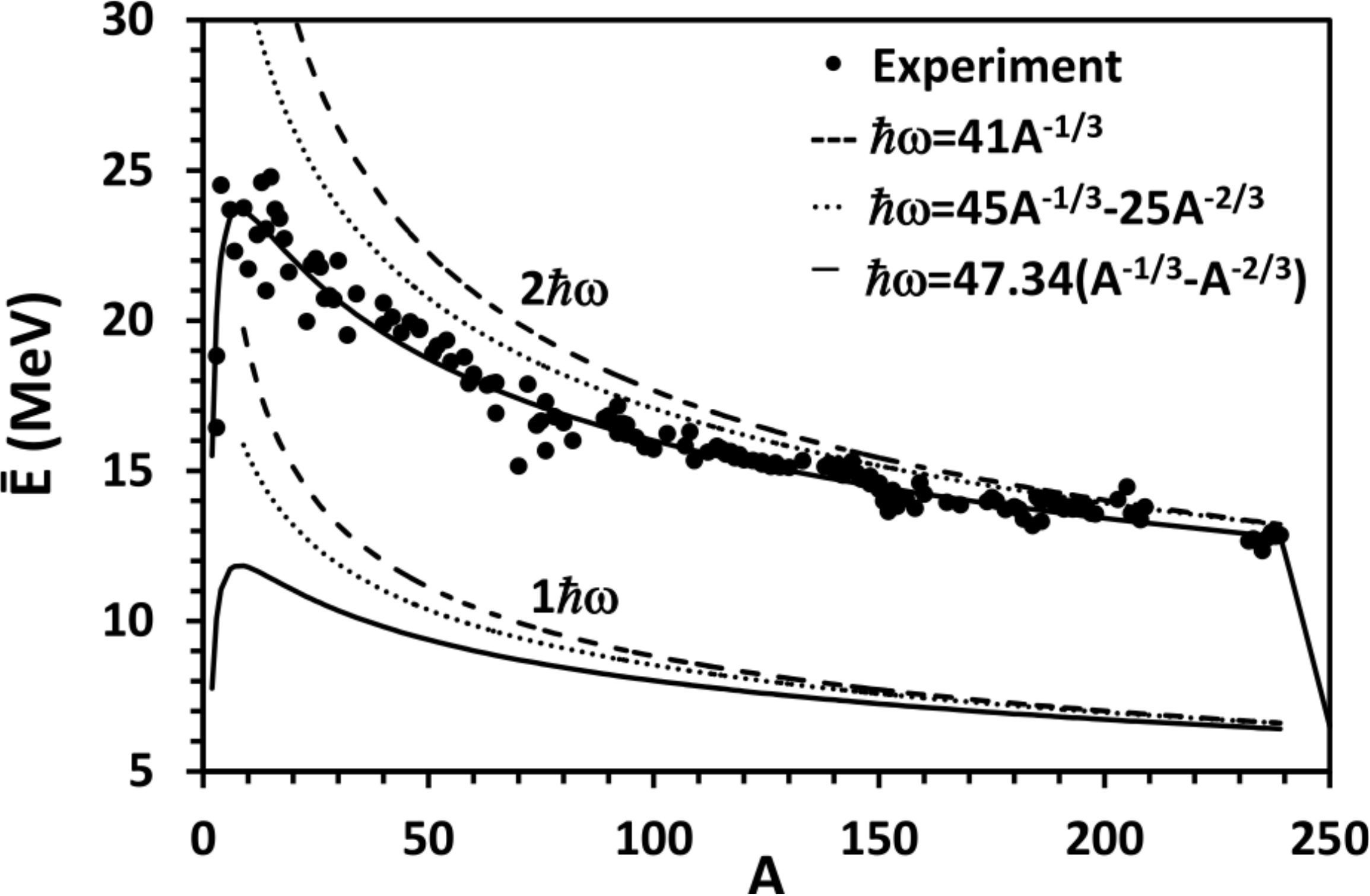}
\caption{The mean experimental GDR energies, $\overline{E}$, defined by Eq.~\ref{EQ3} ($\bullet$) are compared with $\hbar\omega$ harmonic oscillator gaps calculated by Moszkowski (\textbf{-~-~-}) and by Blomqvist and Molinari (\textbf{$\cdot\cdot\cdot$}).  A least square fit to the mean energies gives $\overline{E}=47.34(A^{-1/3}-A^{-2/3})$ (\textbf{---}).}
  \label{E-GDR}
\end{figure}

The mean GDR energy, $\overline{E}$, is defined in Eq.~\ref{EQ3}.
\begin{equation}
\overline{E} =\frac{E_1 \sigma_1 + E_2 \sigma_2}{\sigma_1 + \sigma_2}
\label{EQ3}
\end{equation}
Experimental values of $\overline{E}$ for 148 nuclei~\cite{Kawano19,Varlamov99} are plotted as a function of mass number, $A$, in Fig.~\ref{E-GDR}.  They are compared with the harmonic oscillator energies calculated by Moszkowski~\cite{Mosz57} as given in Eq.~\ref{Mosz} where $r_0$=1.2 fm and
\begin{equation}\label{Mosz}
\hbar\omega \approx \frac{5}{4}\bigg(\frac{\hbar^2}{M_nr_0^2}\bigg)\bigg(\frac{3}{2}\bigg)^{1/3}A^{1/3} = 41A^{-1/3}~\textrm{MeV}
\end{equation}
$M_n$=931.5 MeV is the nucleon mass, and from Blomqvist and Molinari~\cite{Blomqvist68} derived from nuclear charge radii as given in Eq.~\ref{BM}.  A least squares fit to the GDR energies
\begin{equation}\label{BM}
\hbar\omega\approx 45A^{-1/3}-25A^{-2/3}~\textrm{MeV.}
\end{equation}
of nuclei with  ${A=2-239}$ is given by Eq.~\ref{EQ4}.  Remarkably
\begin{equation}
\label{EQ4}
\overline{E}= 2\hbar\omega = 2(47.34\pm0.27)(A^{-1/3}-A^{-2/3})
\end{equation}
the GDR energy can be fit with a single parameter to better than 0.6\% for all nuclei, with a coefficient of determination~\cite{COD} $R^2$=1.00.  This fit even accounts for the rapid decrease in GDR energy for $^3$He~\cite{Faul81} and $^6$Li~\cite{Dytlewski84}.

The mean GDR energies are systematically lower than the harmonic oscillator energies calculated by Moszkowski and by Blomqvist and Molinari for $A\gtrsim100$.  This shift can be ascribed to the $E_{\gamma}^{-3}$ dependence of the reduced widths of dipole transitions.  As shown in Fig.~\ref{GDRHO} removing the energy dependence shifts the photon strength peak by $\approx$1 MeV comparable to the difference between the GDR energy and the $2\hbar\omega$ harmonic oscillator energy.  Below $A=100$ the mean GDR energies vary rapidly from the calculated harmonic oscillator energies which appear to become divergent at low masses.

At each harmonic oscillator energy gap a new ensemble of shell model states becomes available whose wavefunctions mix with those of lower states to generate an increase in level density.  The GDR is thus consistent with this increase in level density and no collective enhancement of photon strength is required.

\begin{figure}[!ht]
\centering
    \includegraphics[width=8cm]{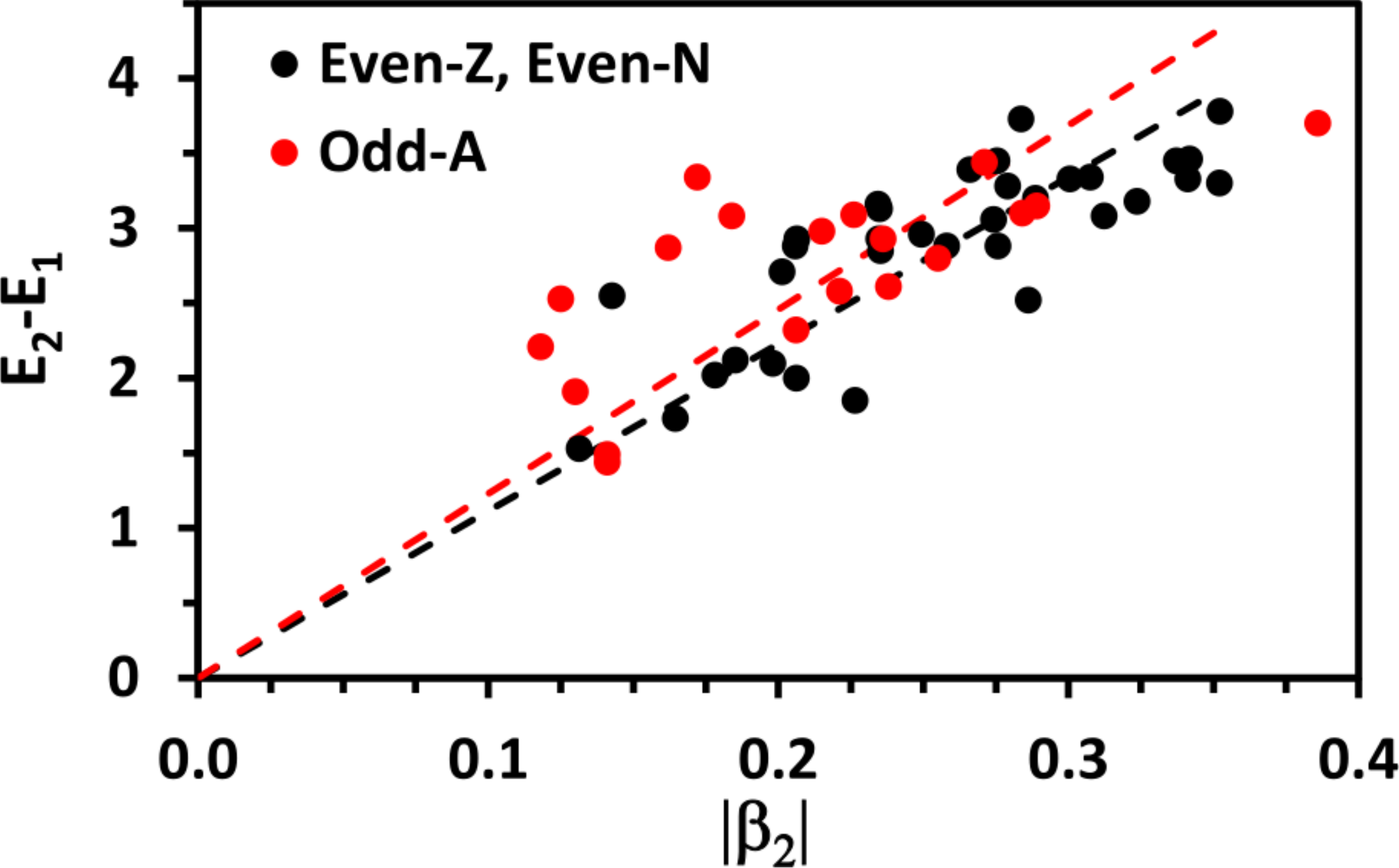}
\caption{Experimental GDR peak separation, ${\Delta E = E_2-E_1}$,  plotted using experimental ($\bullet$) and theoretical (\textcolor{red}{$\bullet$}) $|\beta_2|$ values.}
\label{DE-GDR}
\end{figure}

The GDR separates into two peaks in deformed nuclei.  The measured energy separations, ${\Delta E = E_2-E_1}$ for 37 even-even nuclei and 19 odd-A nuclei are plotted versus the absolute values of their ground state $|\beta_2|$ deformations in Fig.~\ref{DE-GDR}.  The $\beta_2$ parameters were calculated from the experimental B(E2) values~\cite{Pritychenko17} for even-even nuclei, using Eq.~\ref{EQ5}, and taken from the theoretical values of
\begin{equation}
\beta_2 = B(E2)^{1/2}\frac{4\pi}{3Zr_0^2}
\label{EQ5}
\end{equation}
Möller et al~\cite{Moller16} for odd-A nuclei.  Although there is considerable scatter in the data, $\Delta E$ varies linearly with deformation.  The even-even and odd-A data are least-squares fit as shown in Eq.~\ref{Beta} with ${R^2=0.98}$ for even-even
\begin{equation}\label{Beta}
\begin{aligned}
\Delta E =& (11.1\pm0.3)|\beta_2|~\textrm{Even-Z,Even-N}\\
\Delta E =& (12.3\pm0.6)|\beta_2|~\textrm{Odd-A, Odd-Z,Odd-N}
\end{aligned}
\end{equation}
nuclei and ${R^2=0.95}$ for odd-A nuclei.  No doublet GDR measurements were reported for nuclei with $\beta_2<0.12$ most likely because these cases were unresolved experimentally.  The consistent fit using both experimental and theoretical $\beta_2$ indicates that the theoretical values are adequate for this analysis although experimental values should be preferred  when available.

\subsection{GDR cross section}

\begin{figure}[!ht]
\centering
    \includegraphics[width=8cm]{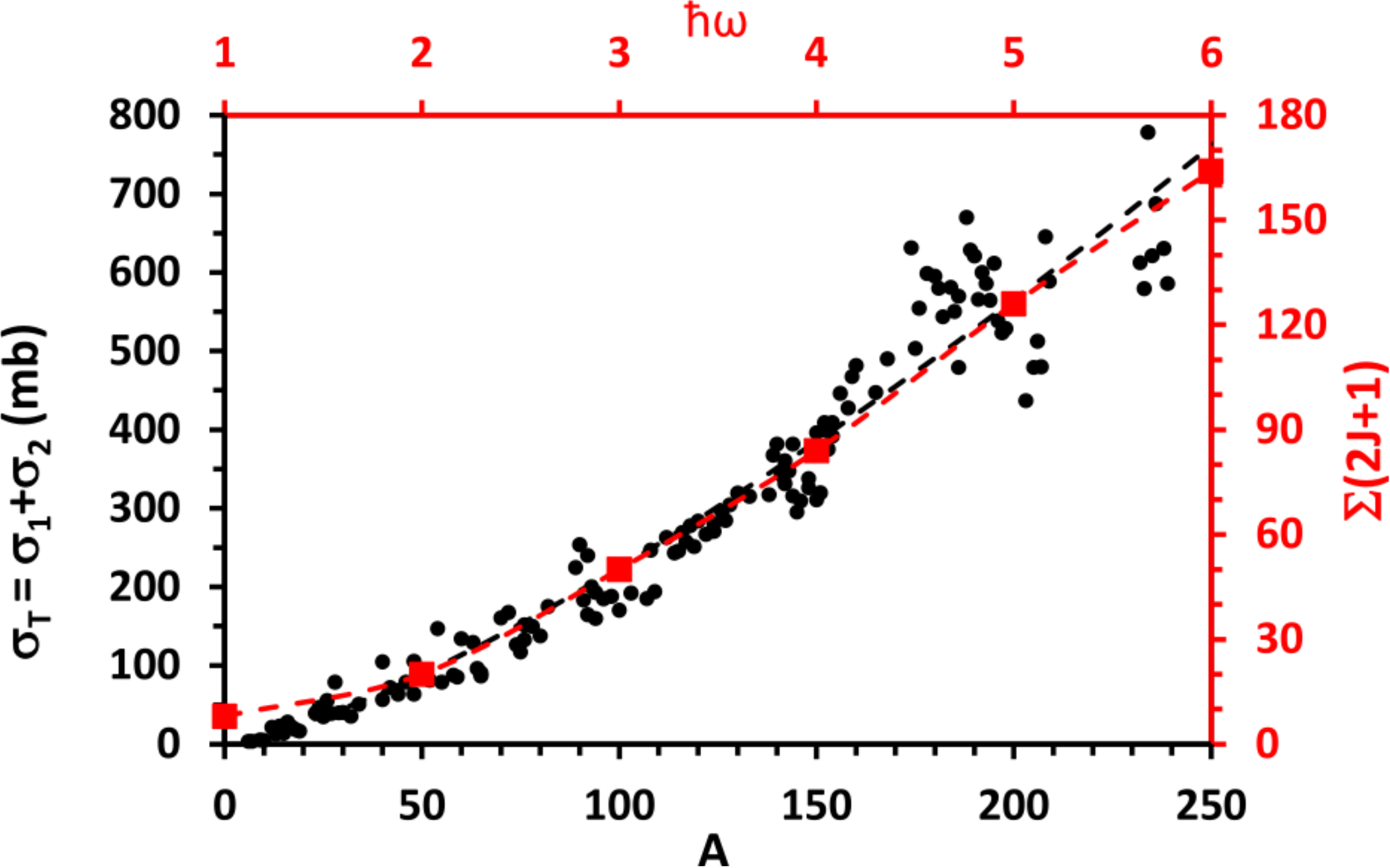}
\caption{The total GDR cross section ${\sigma=\sigma_1+\sigma_2}$ ($\bullet$), least squares fit to $A^{4/3}$ (\textbf{---}), is compared to the increase in magnetic substates, $\Sigma(2J+1)$, (\textcolor{red}{\textbf{$\blacksquare$}}) summed over all Nilsson states occurring at  each shell gap, $\hbar\omega$.}
\label{S-GDR}
\end{figure}

The total measured GDR cross sections, ${\sigma_T=\sigma_1 + \sigma_2}$, are plotted versus mass in Fig.~\ref{S-GDR}.  The total cross section varies proportionally to $A^{4/3}$ as given in Eq.~\ref{SIGF} where a
\begin{equation}\label{SIGF}
\sigma_T = (0.483\pm0.005)A^{4/3}
\end{equation}
least-squares fit gives a ${R^2=0.98}$.  This fit is not unique as a fit to atomic number gives $\sigma_T = (0.794\pm0.009)Z^{3/2}$ and a fit to neutron number gives $\sigma_T = (0.979\pm0.011)N^{4/3}$ both with ${R^2=0.98}$.

\begin{figure}
  \centering
  \includegraphics[width=8.5cm]{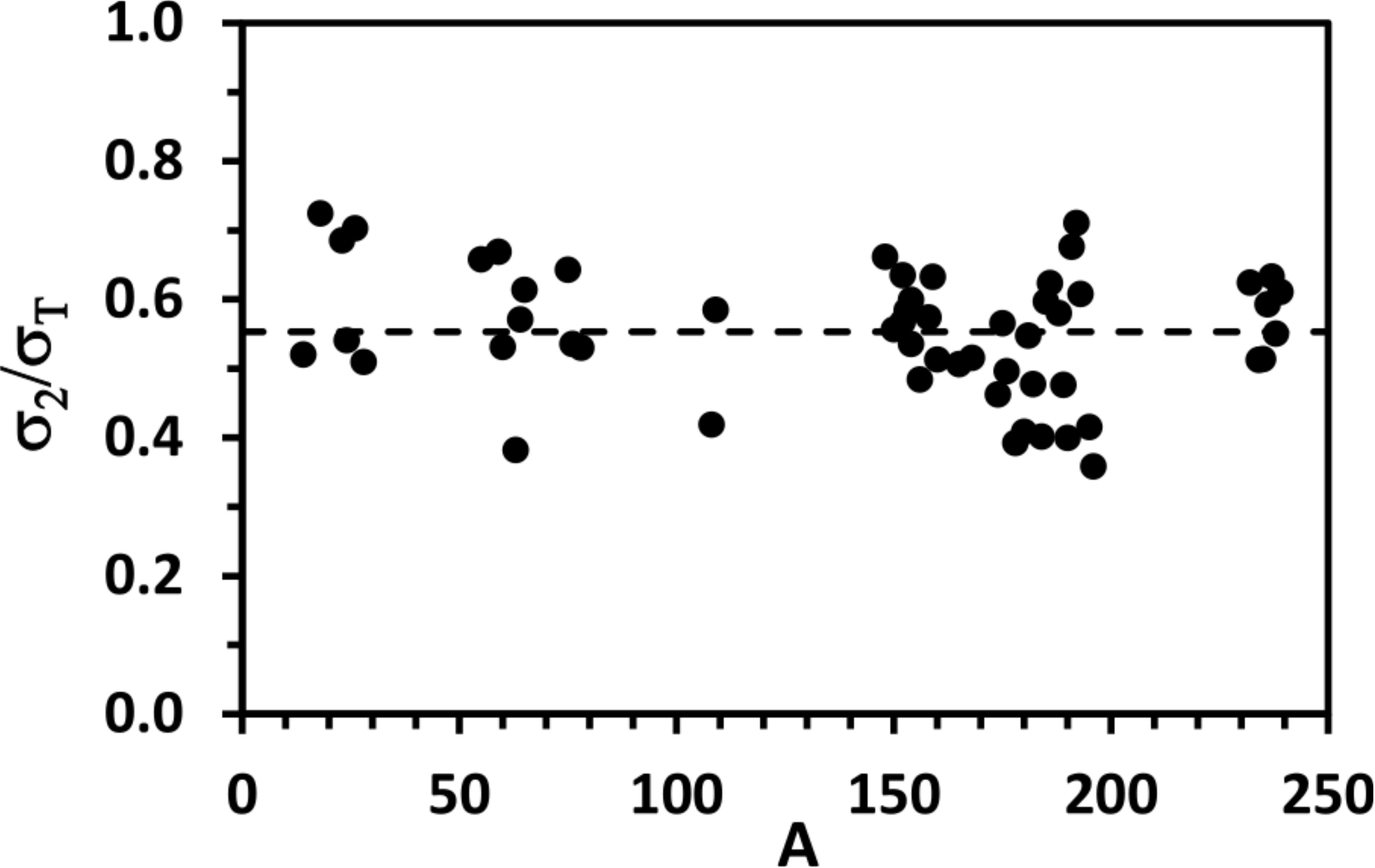}
  \caption{Variation of the ratio $\Gamma_2/\Gamma_T$ ($\bullet$) with mass $A$.  A least-squares fit gives ${\Gamma_2/\Gamma_T=0.55\pm0.09}$ (-~-~-).  Eight outlier data with $\Gamma_2/\Gamma_T>0.8$ and $\Gamma_2/\Gamma_T<0.2$ have been excluded.}\label{SIG2F}
\end{figure}

For deformed nuclei the experimental ratios $\sigma_2/\sigma_T$ are plotted in Fig.~\ref{SIG2F}.  The data scatter significantly with an average of $\sigma_2/\sigma_T=0.55\pm0.09$.  The scatter in the data is likely an experimental artifact due to the difficulty in resolving the two GDR peaks.  If the total cross section varies smoothly with $A^{4/3}$, then $\sigma_2/\sigma_T$ should also vary smoothly.

The increase in the GDR cross section with mass correlates with an increase in the number of Nilsson configuration magnetic substates occurring at each oscillator gap as shown in Fig.~\ref{S-GDR}.  In the Nilsson model each shell model state evolves into series of collective configurations $\Omega[Nn_Z \Lambda]$ each contributing $2\Omega+1$ magnetic substates~\cite{Firestone96}.  At $1\hbar\omega$ only 8 Nilsson configurations emerge with $\Omega$=1/2,3/2, and at $6\hbar\omega$ 164 configurations occur with $\Omega$=1/2-13/2.  Notably as the harmonic oscillator model evolves into the shell model and then into the Nilsson model the average configuration energy remains constant at the harmonic oscillator energy, $n\hbar\omega$.  The increase in Nilsson configuration magnetic substates is a surrogate for the increase in total level density so the GDR cross section is proportional to the level density which increases with an $A^{4/3}$ dependence.  Experimental GDR cross sections vary from Eq.~\ref{SIGF} by $<$25\% in all cases.

\subsection{GDR width}

\begin{figure}[!ht]
\centering
    \includegraphics[width=8cm]{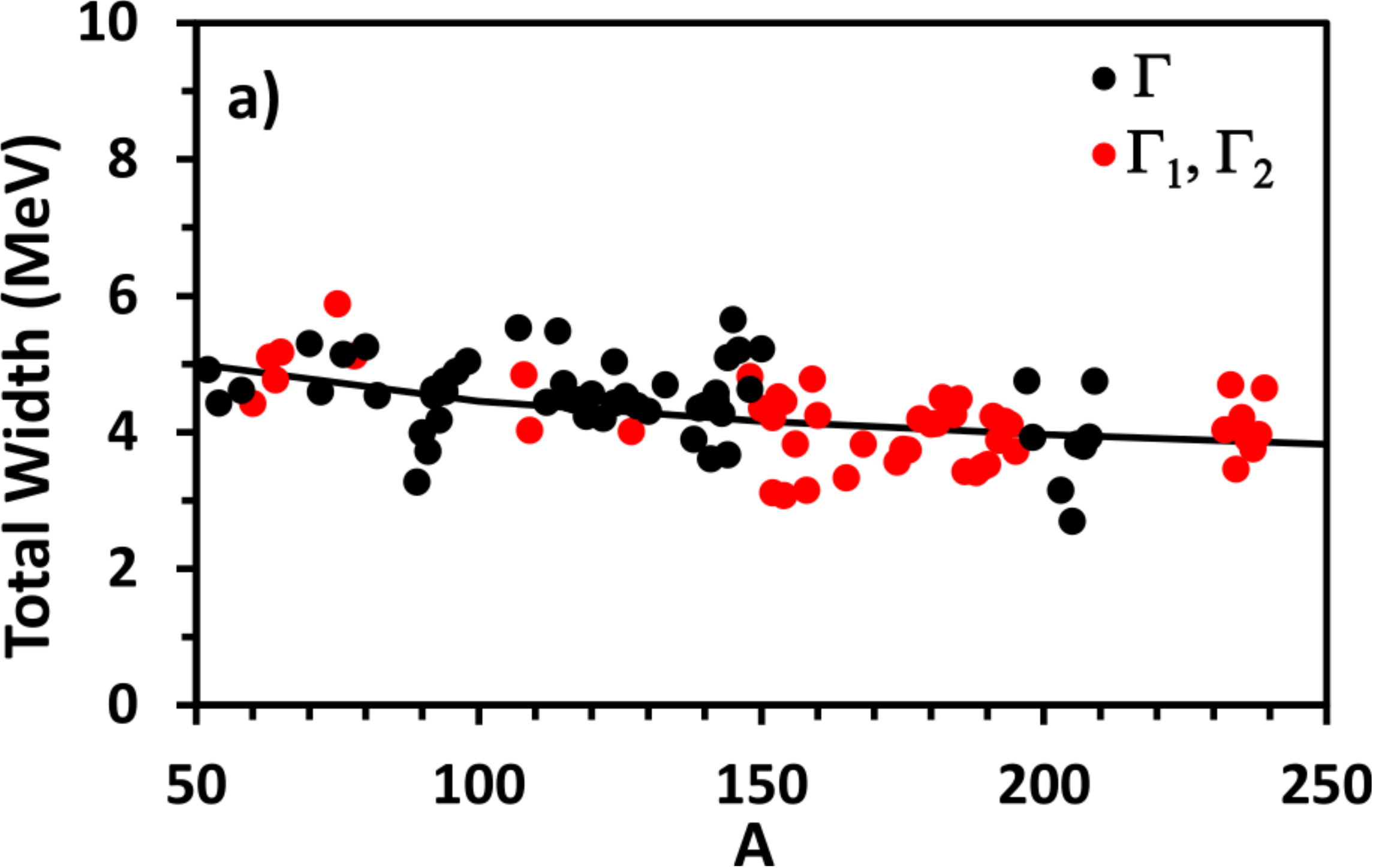}
    \includegraphics[width=8cm]{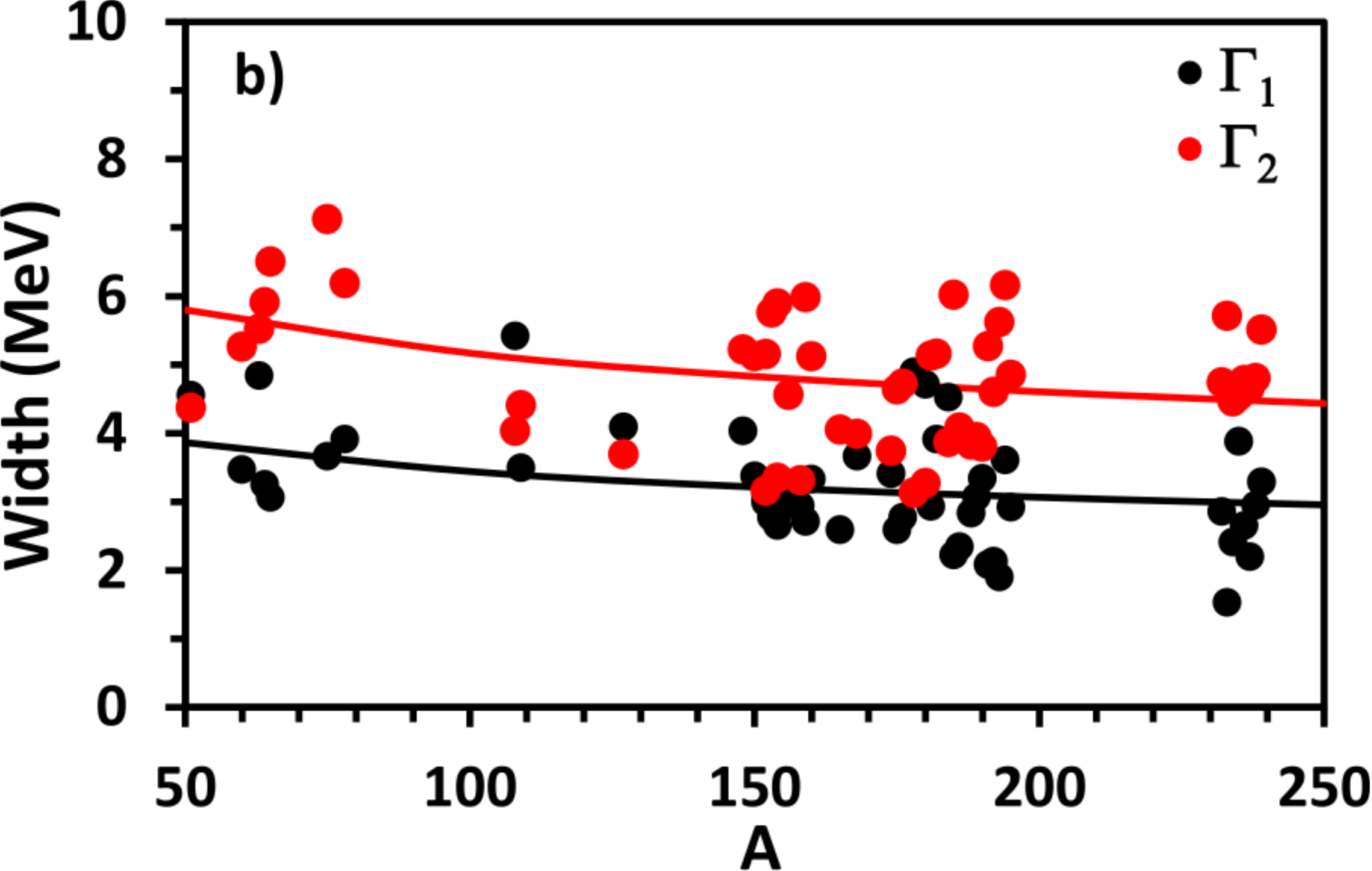}
\caption{a) Total widths $\Gamma$ for nuclei with a single peak width and ${\Gamma_{1,2}=(\Gamma_1\sigma_1+\Gamma_2\sigma_2)/(\sigma_1+\sigma_2)}$ where separate peak widths were measured.  b) Individual widths $\Gamma_1$ and $\Gamma_2$ where both peak widths were measured.}
\label{W-GDR}
\end{figure}

The width of the GDR corresponds to a Lorentzian distribution of nuclear levels about the GDR peak.  This definition is complicated by the fact that some experiments measure a single GDR peak while others measure two GDR peaks.  The singlet GDR peak widths can be interpreted as experimentally unresolved doublets.  For comparison with singlet GDR peaks a cross section weighted average doublet peak width, $\overline{\Gamma}$, is defined by Eq.~\ref{EQ7}.  The singlet, $\Gamma$, and weighted average doublet, $\overline{\Gamma}$,
\begin{equation}
\overline{\Gamma} = \frac{\Gamma_1 \sigma_1 +\Gamma_2 \sigma_2}{\sigma_1+\sigma_2}
\label{EQ7}
\end{equation}
peak widths are compared as a function of mass, $A$, in Fig.~\ref{W-GDR}a.  Both follow the same narrow mass distribution.  From Eq.~\ref{EQ2} it is clear that asymptotically as $E_{\gamma}\rightarrow0$, $\Gamma\varpropto E_i^{1/2}$.  The $\Gamma$ and $\overline{\Gamma}$ widths are least-squares fit to $\overline{E}^{1/2}$ in Eq.~\ref{GG2} with ${R^2=0.98}$ in both cases.
\begin{equation}\label{GG2}
\begin{aligned}
\Gamma = &(1.118\pm0.0021)\overline{E}^{1/2}\\
\overline{\Gamma} = &(1.098\pm0.0023)\overline{E}^{1/2}\\
\textrm{Average} (\overline{\Gamma},\Gamma) = &(1.110\pm0.0015)\overline{E}^{1/2}
\end{aligned}
\end{equation}

The doublet GDR peak widths, $\Gamma_1$ and $\Gamma_2$ are plotted separately versus mass, $A$, in Fig.~\ref{W-GDR}b.   There is considerable scatter in the data yet the two widths fall into separate groups which are least squares fit as given in Eq.~\ref{G1G2}.  The ratio of the two widths is ${\Gamma_2/\Gamma_1=1.50\pm0.07}$.
\begin{equation}\label{G1G2}
\begin{aligned}
\Gamma_1 =& (0.84\pm0.03)\overline{E}^{1/2}\\
\Gamma_2 =& (1.26\pm0.04)\overline{E}^{1/2}
\end{aligned}
\end{equation}

The GDR peak width ratio, ${\Gamma_2/\Gamma_1+\Gamma_2=0.60\pm0.03}$, is consistent with ${\sigma_2/\sigma_1+\sigma_2=0.55\pm0.09}$ suggesting that the cross section is proportional to the width.  Adopting the more precise ratio, ${\sigma_2/\sigma_1+\sigma_2=0.60\pm0.03}$, and using Eq.~\ref{SIGF} we get the values of $\sigma_1$ and $\sigma_2$ given in Eq.~\ref{S12}.  Combining the
\begin{equation}\label{S12}
\begin{aligned}
\sigma_1 =&(0.193\pm0.010)A^{4/3}\\
\sigma_2 =&(0.290\pm0.015)A^{4/3}
\end{aligned}
\end{equation}
cross section ratio with Eq.~\ref{EQ4} and Eq.~\ref{Beta} gives the individual GDR peak energies as shown in Eq.~\ref{E12}.  This gives
\begin{equation}\label{E12}
\begin{aligned}
E_1 =& \overline{E}-(6.68\pm0.16)|\beta_2|\\
E_2 =& \overline{E}+(4.45\pm0.11)|\beta_2|
\end{aligned}
\end{equation}
a complete description of the GDR photon strength as a function of $A$ and $\beta_2$.

\section{Discussion}

The collective explanation for the GDR is unnecessary because the increase in level density at the shell closure is a much simpler argument.  The BA photon strength parameters describing the GDR are based on fundamental, universal functions of mass and deformation.

\vspace{-0.5cm}

\subsection{GDR and nuclear radii}

\begin{figure}
  \centering
  \includegraphics[width=8.5cm]{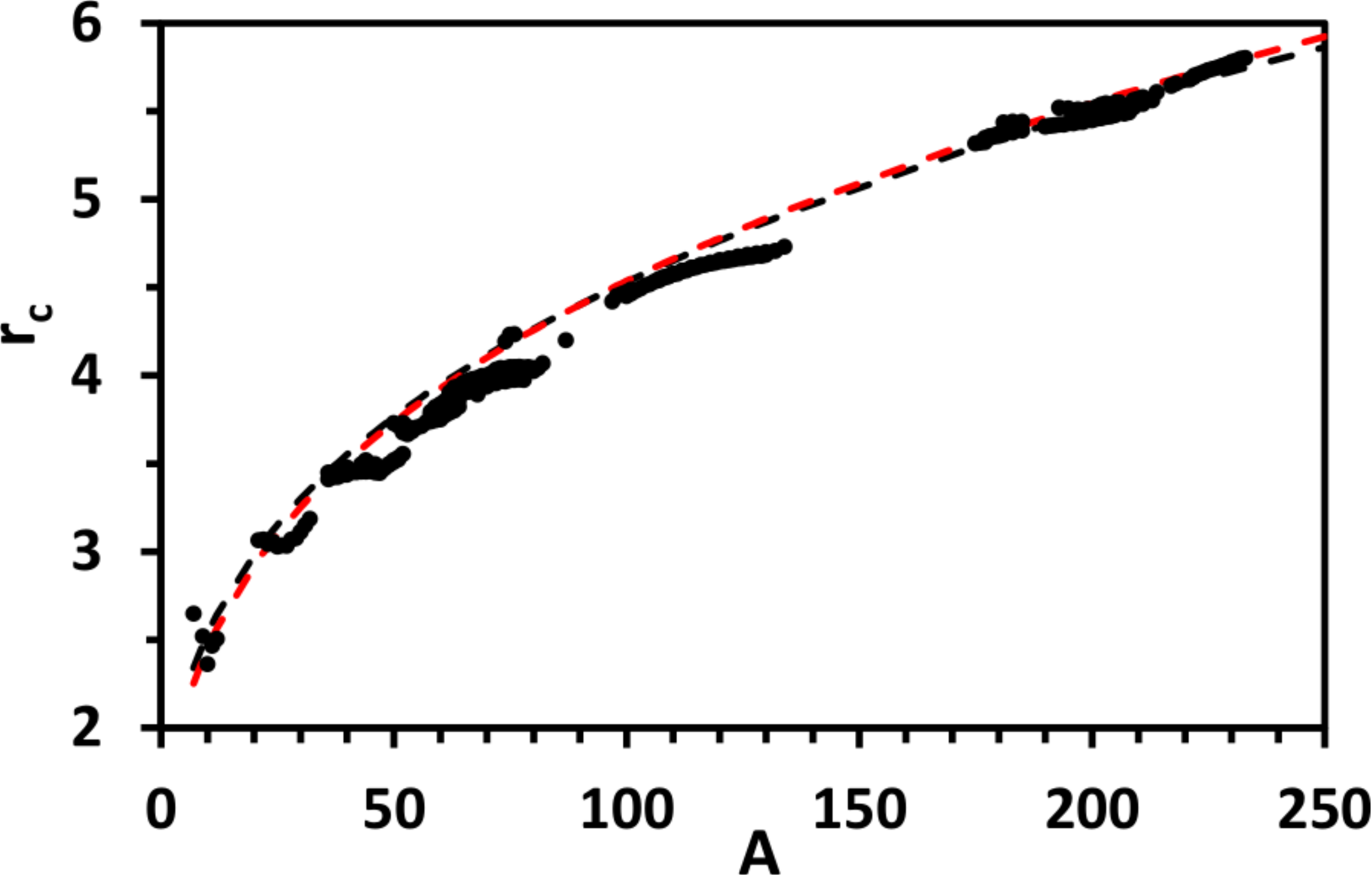}
  \caption{Comparison of experimental nuclear radii ($\bullet$)~\cite{Li21} with values calculated by Blomqvist and Molinari ($\textcolor{red}{\textbf{-~-~-}}$)~\cite{Blomqvist68} and derived from the GDR parameters (\textbf{-~-~-}).}\label{b2P}
\end{figure}

The mean energy of the GDR, $\overline{E}$, is accurately predicted as a function of $A^{-1/3}-A^{-2/3}$.  This functional form is nearly identical to that describing of the harmonic oscillator energy, $\hbar\omega$, by Blomqvist and Molinari~\cite{Blomqvist68} based on the systematics of nuclear radii. The root mean squared radius, $r_c$ is given by Eq.~\ref{MSR} where $b$ is the harmonic oscillator length parameter, $M_n$ is the nucleon
\begin{equation}\label{MSR}
\begin{aligned}
r_c &= b^2r_0\\
b^2 &= \frac{\hbar^2}{M_n(\hbar\omega)}
\end{aligned}
\end{equation}
mass, and $\omega$ is the oscillator frequency.  The parameter $r_0$ is a constant fit to the experimental data.  The values of $b^2$ derived from the nuclear radii and from the mean GDR energy are given in Eq.~\ref{HOB2}.  Recommended experimental
\begin{equation}\label{HOB2}
\begin{aligned}
\textrm{Blomqvist and Molinari:~}~b^2 =& 0.90A^{1/3}+0.70~\textrm{fm}\\
\textrm{Giant Dipole Resonance:~}~b^2 =& 0.88A^{1/3}+0.88~\textrm{fm}
\end{aligned}
\end{equation}
root mean squared atomic radii~\cite{Li21} are compared with calculated values from Blomqvist and Molinari, assuming $r_0$=0.93, and from the GDR parameters, assuming $r_0$=0.91, in Fig.~\ref{b2P}.  The agreement is excellent with $R^2$=1.00 in both cases.

\subsection{GDR and the Nilsson model}

\begin{figure}[!ht]
\centering
    \includegraphics[width=8cm]{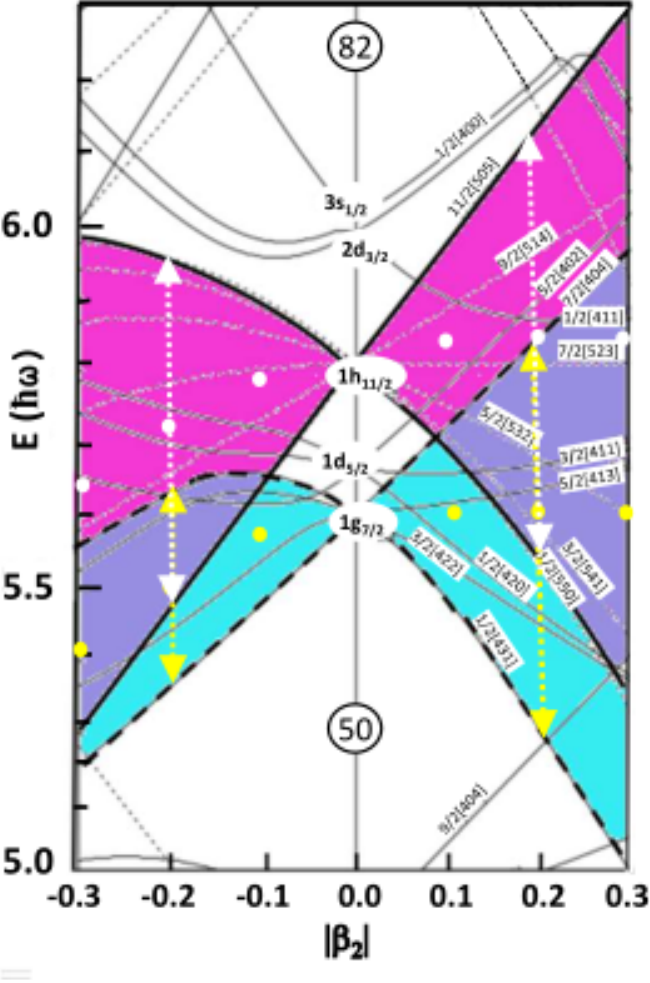}
\caption{Nilsson diagram for the Z=50-82 closed shells region~\cite{Firestone96}.  The region encompassing Nilsson configurations originating from the $1g_{7/2}$ shell model state are bordered by a dashed line and shaded in blue, the region of configurations originating from the $1h_{11/2}$ shell model state are bordered by a solid line and shaded in pink. The shaded region also overlaps the $1d_{5/2}$ configurations.  The centroid energies are indicated for the $1g_{7/2}$ Nilsson states (\textcolor{yellow}{$\bullet$}) and for $1h_{11/2}$ Nilsson states (o).  The yellow and white dotted lines show the energy ranges of $1g_{7/2}$ and $1h_{11/2}$ Nilsson states, respectively at $|\beta_2|$=0.2.}
\label{Nilsson}
\end{figure}

The separation of the GDR into two peaks with different widths is qualitatively described by the Nilsson deformed shell model as shown in the Nilsson diagram for the Z=50-82 region in Fig.~\ref{Nilsson}~\cite{Firestone96}.  In this calculation prolate states arising from the shell model configurations are more widely spaced in energy than oblate states with the same deformation.  For $|\beta_2|$=0.2 the ratio of the ranges of energy spacings of the rotational configurations arising from the $1h_{11/2}$, $1g_{7/2}$, and $1d_{5/2}$ shell model states for prolate and oblate deformations is ${\Delta E(prolate)/\Delta E(oblate)\approx1.6}$ in good agreement with the observed $\Gamma_2/\Gamma_1$ ratio.  The mean energies of the Nilsson prolate configurations are also higher than those for oblate configurations, consistent with the lower GDR peak consisting of transitions to states with oblate deformation and the upper GDR peak to states with prolate deformation.  This indicates that photonuclear reactions populate both oblate and prolate deformed states having the same deformation as the GS.

The cross section populating the higher, prolate GDR peak is 50\% larger than that populating the lower oblate peak despite the expected $E_{\gamma}^{-3}$ energy dependence.  Most deformed nuclei have prolate ground state shapes so this would be consistent with an expected hindrance of prolate to oblate transitions.  For oblate ground states we might expect the cross section populating the lower energy oblate GDR peak to be higher.  There are too few measurements on oblate nuclei to adequately test this hypothesis.

\begin{figure*}[!ht]
  \centering
  \includegraphics[width=8.5cm]{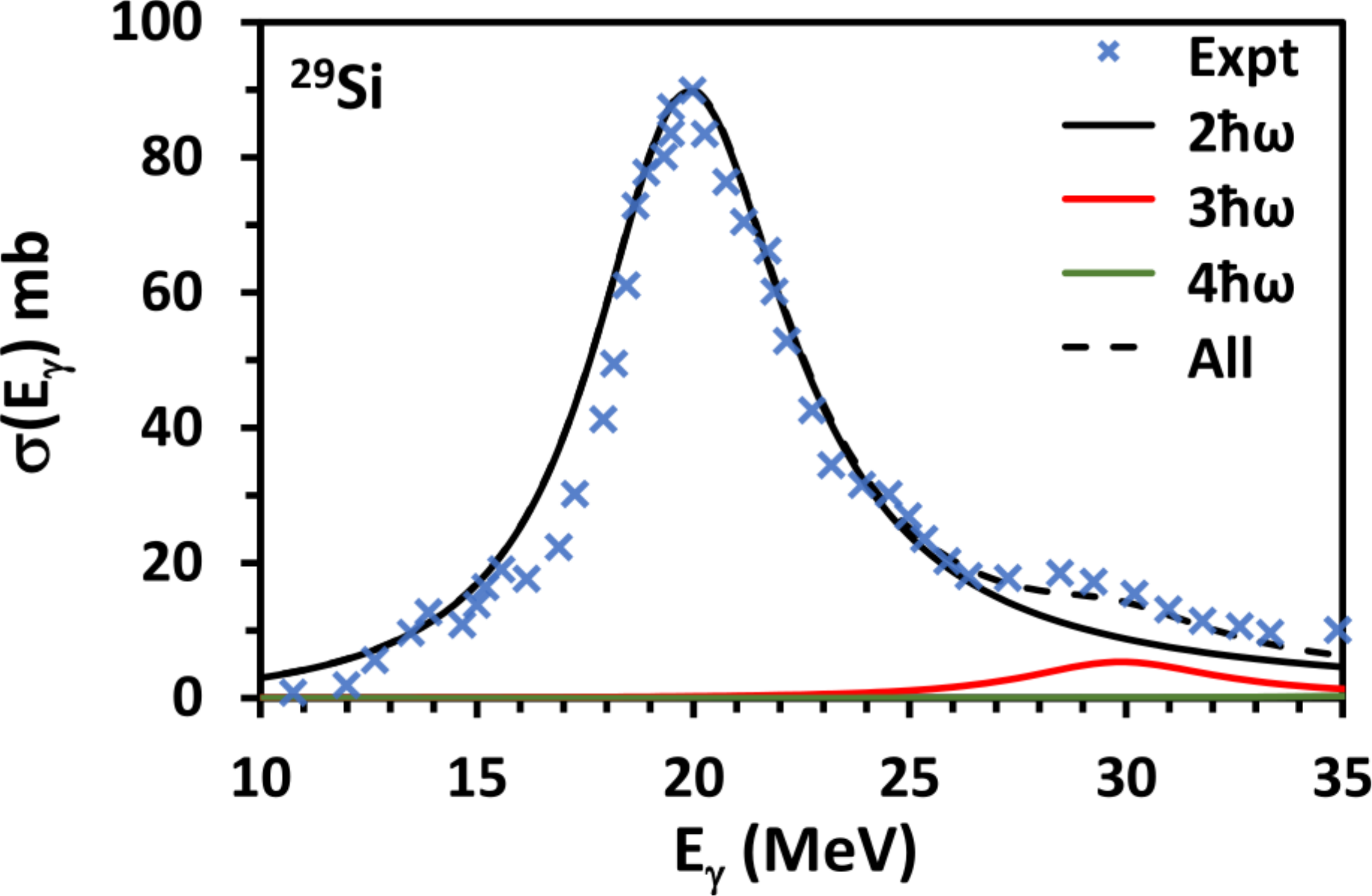}
  \includegraphics[width=8.5cm]{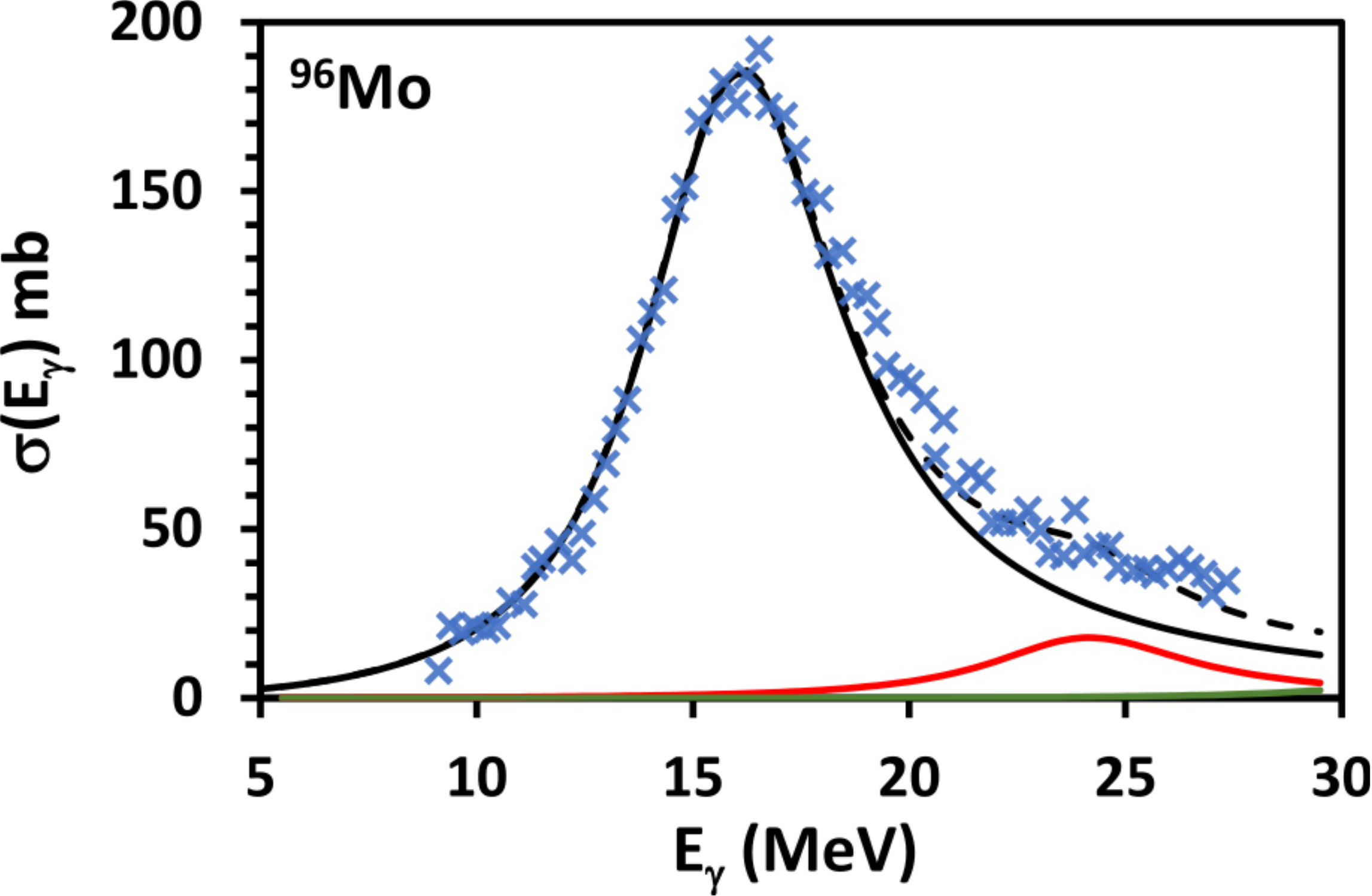}
  \includegraphics[width=8.5cm]{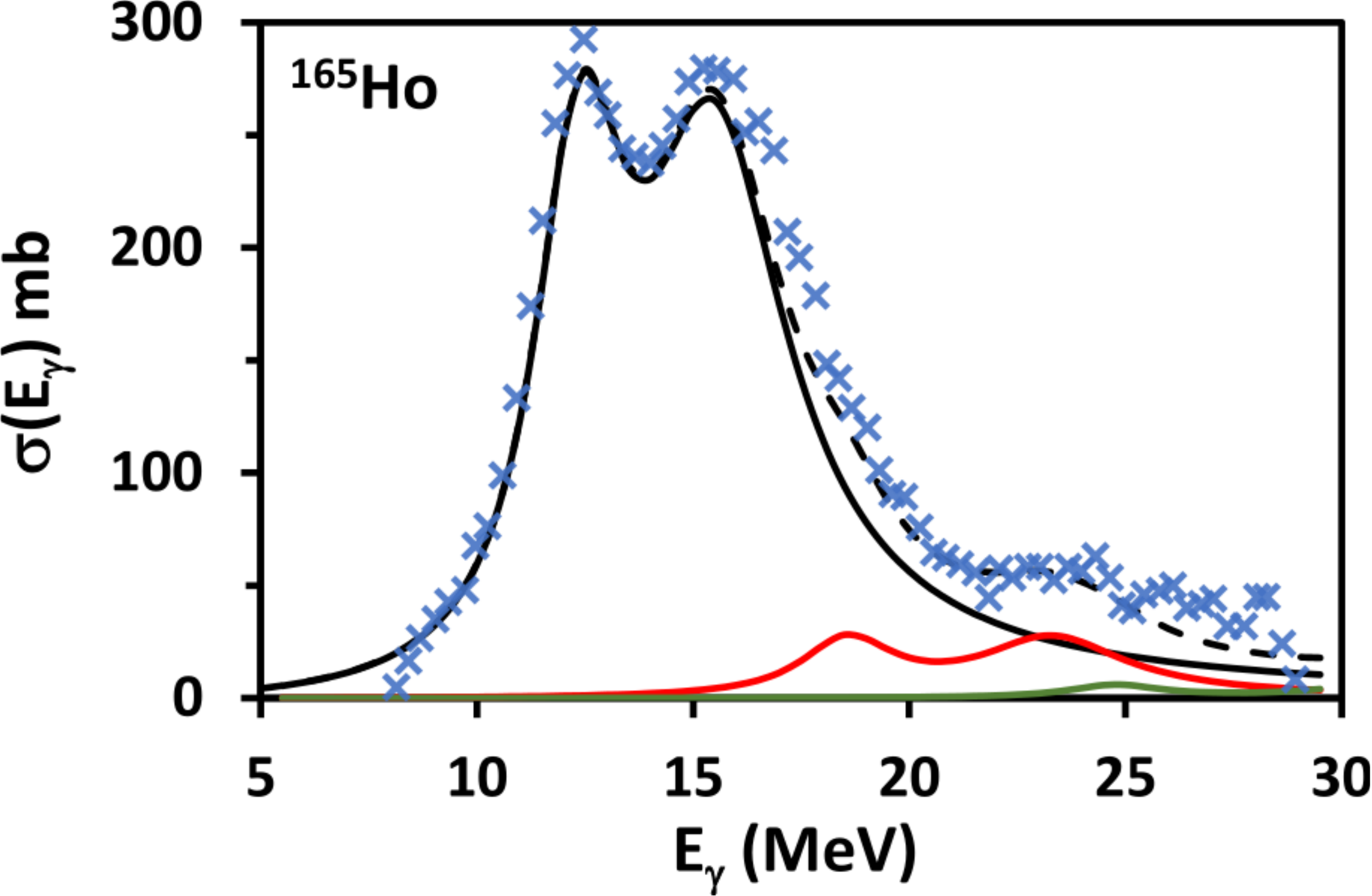}
  \includegraphics[width=8.5cm]{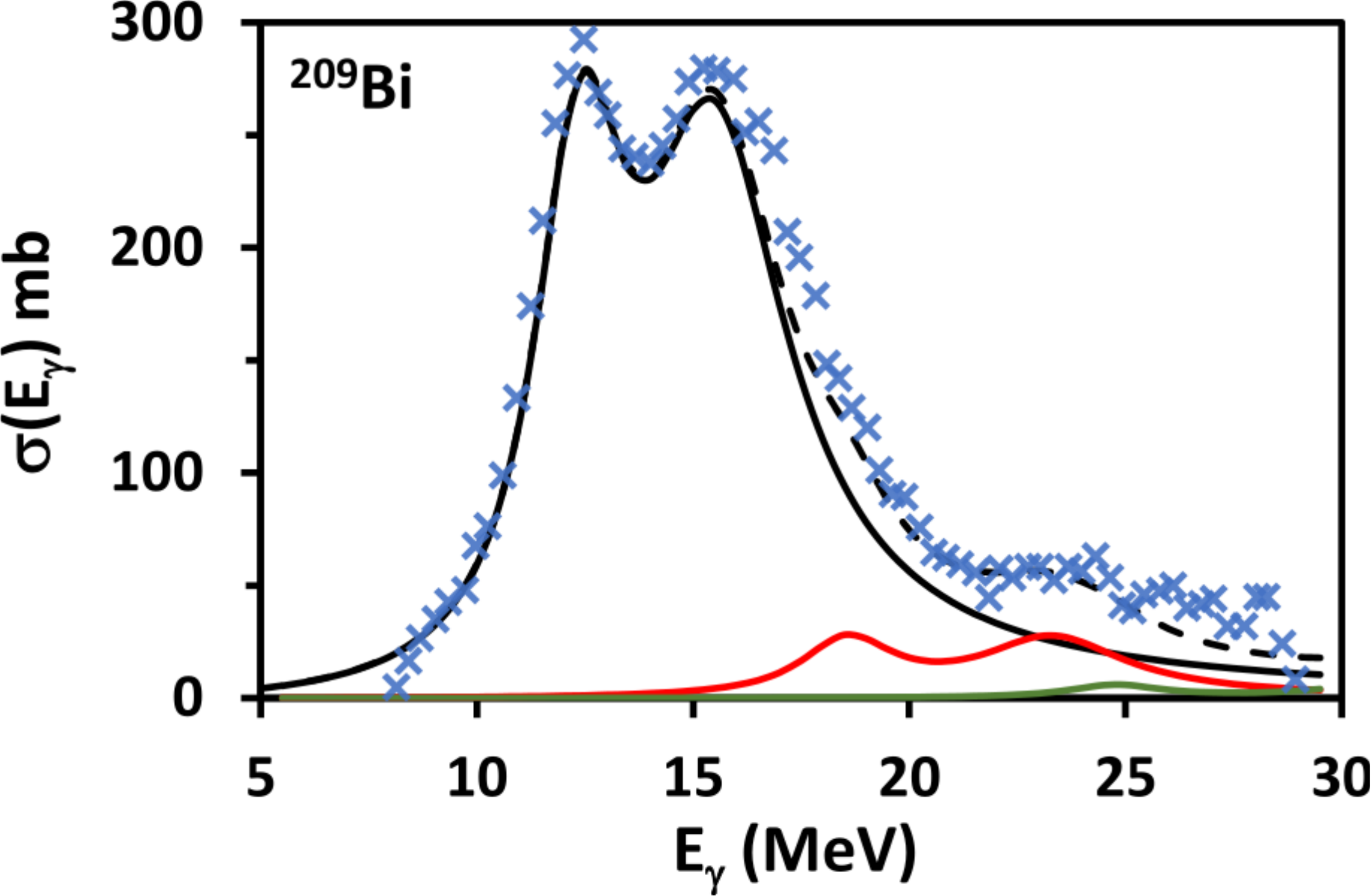}
  \caption{Comparison of experimental GDR cross sections (\textcolor{blue}{{\bf x}}) with calculated values for the $2\hbar\omega$ ({\bf --}), $3\hbar\omega$ (\textcolor{red}{{\bf --}}), $4\hbar\omega$ (\textcolor{calpolypomonagreen}{{\bf --}}) and for all higher resonances ({\bf -~-}).}
  \label{GDR2}.
\end{figure*}

\subsection{Higher order GDR peaks}

If the GDR arises from an increase in level density at an excitation of $2\hbar\omega$ then similar increases must occur at all higher oscillator gap energies.  The higher order GDR peaks will occur at the well known energies, $\overline{E}(n\hbar\omega)$ as given by Eq.~\ref{EQ4}.  At each oscillator energy gap the dependence on deformation, $|\beta_2|$ is likely to be similar because it is determined primarily by the GS deformation.  A similar $\Gamma_2/\Gamma_1$ ratio is also predicted from the Nilsson model.  The higher order GDR cross sections can be predicted from detailed balance where the cross section is proportional to photon strength as shown in Eq.~\ref{SNX} where $f_{n\hbar\omega}$
\begin{equation}
\label{SNX}
\sigma(n\hbar\omega) = \sigma_{2\hbar\omega} \frac{f_{n\hbar\omega} E_{n\hbar\omega}} {f_{2\hbar\omega} E_{2\hbar\omega}}
\end{equation}
is the BA photon strength at each oscillator energy gap.

The experimental GDR cross sections for $^{29}$Si, $^{96}$Mo, $^{165}$Ho, and $^{209}$Bi~\cite{Varlamov99} are compared with calculated values using the IAEA Photonuclear Data Library GDR parameters~\cite{Kawano19} in Fig.~\ref{GDR2}.  In each case the cross section at energies above the GDR peak energy are increased with respect to the standard BA calculation, mainly due to the contribution of the $3\hbar\omega$ resonance.  Since this contribution was ignored in the evaluation of the IAEA library the GDR parameters they may have been fitted incorrectly.  In some cases the excess high energy cross section may have been ascribed to the onset of addition reaction channels, e.g. $(\gamma,2n)$, even though those channels are unlikely to become significant until a few MeV above their threshold energy.
\vspace{-0.5cm}
\subsection{Other giant resonances}

Many other giant resonances populated by nuclear reactions have been reported including the E0 Giant Monopole Resonance (GMR), E1 Isoscalar Giant Dipole Resonance (ISGDR), E2 Giant Quadrupole Resonance (GQR), and the E3 Giant Octupole Resonance (GOR).  It is likely that all of these giant resonances are associated with an increase in level density at a harmonic oscillator energy gap.

\begin{figure}[!ht]
\centering
    \includegraphics[width=8cm]{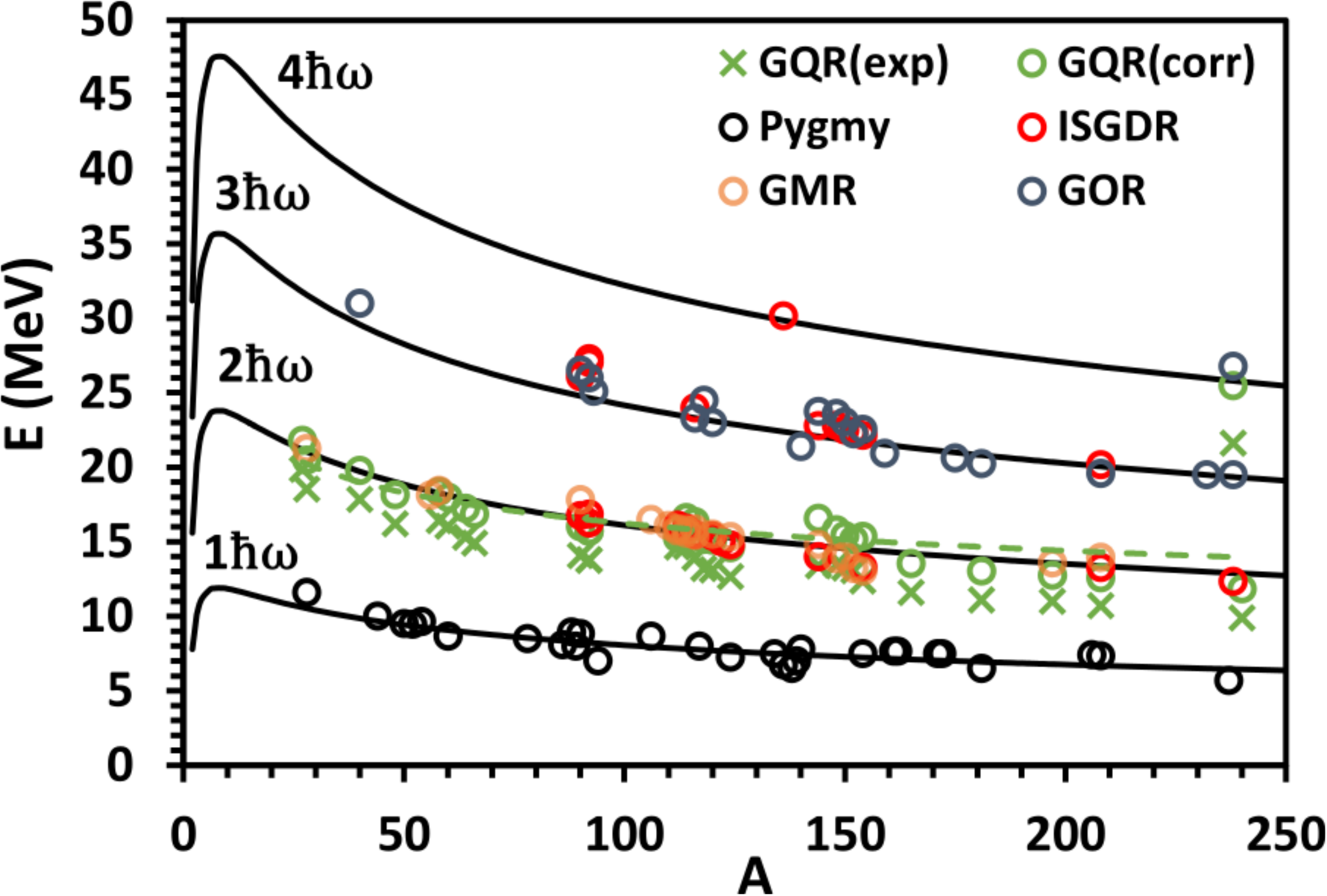}
\caption{The experimental energies of Pigmy and spin flip $(\textbf{o})$, isoscalar giant dipole ISGDR ($\textcolor{red}{\textbf{o}}$), giant quadrupole GQR ($\textcolor{cadmiumgreen}{\textbf{o}}$) giant monopole GMR ($\textcolor{brown}{\textbf{o}}$), and giant octupole GOR ($\textcolor{blue}{\textbf{o}}$) resonances are compared with the GDR harmonic oscillator energies. The experimental GQR energies ($\textcolor{cadmiumgreen}{\textbf{x}}$) are corrected as described in the text.}
\label{GR-all}
\end{figure}

At $1\hbar\omega$ a giant resonance should also occur although none is widely recognized.  This giant resonance can be ascribed to the strong Pygmy (E1) and spin flip (M1) transitions observed in many nuclei.  They tend to be highly fragmented in energy with strong single transitions standing out because of the relatively low level density in this region.  The average experimental energies of pygmy and spin flip transition groups in many nuclei are plotted in   Fig.~\ref{GR-all}~\cite{Flynn74,Richter83,Scheck13,Schramm12,Schweng13,Schweng07,Beno09,Kamer06,Iwam12,Schweng08,Derya13,Erik14,Toft11,Endres10,Mass14,Savran08,Endres09,Makin10,Sarri93,
Voinov01,Lewis72,Polto12,Lasze85,Tornyi14}.  Their energies agree well with the $1\hbar\omega$ oscillator energy gap.

Energies of the (GMR)~\cite{Clark99,Bert79,Lebrun80,Patel12,Young97,Lui84,Lasze13,Sharma88,Young02,Itoh03,Young04,Lui04,Lui06,Monro08,Doer76}, (ISGDR)~\cite{Clark01,Uchi04,Young04,Li10,Gupta18,Schmidt93,Davis97}, (GQR)~\cite{Young81,Szik84,Sharma88,Young02,Itoh03,Young04,Lui04,Lui06,Gupta18,Monro08}, and (GOR)~\cite{Morsch82,Yama81,Saito83,Pitt80,Young04,Carey80,Itoh04} giant resonances are also plotted in Fig.~\ref{GR-all}.  All but the GQR follow the $2\hbar\omega$, $3\hbar\omega$ and $4\hbar\omega$ harmonic oscillator gap energies as predicted.
The experimental GQR resonance energies are systematically 1-3 MeV below the $2\hbar\omega$ oscillator gap energy.  This can be explained by the $1/E_{\gamma}^{5}$ energy dependence of the GQR photon strength which shifts its energy down as shown in Fig.~\ref{GDRHO}.  The higher GQR peak is also strongly suppressed explaining why most measurements of the GQR reported only a single peak.  Itoh \textit{et al}~\cite{Itoh2003} have shown that for the samarium isotopes the GQR is composed of two peaks and that the higher energy peak is significantly weaker.  In Fig.~\ref{GDRHO} the measured $^{238}$U GQR photon strength~\cite{Arruda80} is compared with the BA prediction corrected for a $1/E_{\gamma}^{5}$ energy dependence.  The agreement is excellent except that the experimental data miss the second GQR peak.  Shifting all of the GQR energies by 2 MeV gives an excellent agreement with the $2\hbar\omega$ oscillator gap energy as shown in Fig.~\ref{GR-all}.

The occurrence of all giant resonances at oscillator gap energies confirms that they can all be explained by increases in level density at the shell closures.  No more complex collective symmetry arguments are necessary to explain them.

\section{Conclusions}

The GDR photon strength can be fully described by a Lorentzian distribution whose parameters are defined by the harmonic oscillator and Nilsson deformed shell models.  The energy of the GDR follows the $2\hbar\omega$ oscillator shell gap energy in a formulation nearly identical to that derived from nuclear radii.  The division of the GDR into two peaks for deformed nuclei is completely consistent with the expectations of the Nilsson model.  The association of the GDR with the harmonic oscillator energy gap is fundamental to all giant resonances.  There is no a priori reason to invoke collection motion as the cause of giant resonances when all can be explained by the increase in level density expected at the harmonic oscillator energy gaps.

\begin{table}[!ht]
\tabcolsep=12pt
\caption{\label{Parameters} Giant Dipole Resonance Brink-Axel photon strength parameters.}
\begin{tabular}{clc}
\toprule
&${\overline{E}=2(47.34\pm0.27)(A^{-1/3}-A^{-2/3})}$&\\
&${E_1 = \overline{E} - (6.68\pm0.16) |\beta_2|}$&\\
&${E_2 = \overline{E} + (4.45\pm0.11) |\beta_2|}$&\\
&${\sigma_1 = (0.193\pm0.010)A^{4/3}}$&\\
&${\sigma_2 = (0.290\pm0.015)A^{4/3}}$&\\
&${\Gamma_1 = (0.84\pm0.03)\overline{E}^{1/2}}$&\\
&${\Gamma_2 = (1.26\pm0.04)\overline{E}^{1/2}}$&\\
\botrule
\end{tabular}
\end{table}

The parameters necessary to define the GDR Lorentzian distribution are summarized in Table~\ref{Parameters}.  They are all a function of mass, $A$, and deformation, $|\beta_2|$.  Although deformations are not experimentally known for all nuclei, an adequate source of calculated values is available from Möller et al~\cite{Moller16}.  These parameters were derived from experimental data based on difficult analyses that may have biased their results.  This universal formulation of the GDR has broad applicability and would be an asset in analyzing future measurements.  It has not been explored whether second order effects, especially for nuclei near magic numbers, might be important although these effects should be $<$20\% in the most cases.  The parameterization discussed here is only valid for nuclei near stability as no data for nuclei far from stability is available.

\section{Acknowledgements}

This work was performed with the support from the University of California Retirement system without any US federal funding.  The author wishes to thank the Nuclear Data Section of the International Atomic Energy Agency and the University of Oslo Nuclear Physics group for previous support and discussions that ultimately led to the ideas presented in this paper.  I would also like to thank Vladimir Varlamov, Moscow State University, for his kind comments on this paper.


\begin{thebibliography}{72}
\expandafter\ifx\csname natexlab\endcsname\relax\def\natexlab#1{#1}\fi
\expandafter\ifx\csname bibnamefont\endcsname\relax
  \def\bibnamefont#1{#1}\fi
\expandafter\ifx\csname bibfnamefont\endcsname\relax
  \def\bibfnamefont#1{#1}\fi
\expandafter\ifx\csname citenamefont\endcsname\relax
  \def\citenamefont#1{#1}\fi
\expandafter\ifx\csname url\endcsname\relax
  \def\url#1{\texttt{#1}}\fi
\expandafter\ifx\csname urlprefix\endcsname\relax\def\urlprefix{URL }\fi
\providecommand{\bibinfo}[2]{#2}
\providecommand{\eprint}[2][]{\url{#2}}

\bibitem[{\citenamefont{Goldhaber and Teller}(1948)}]{Goldhaber48}
\bibinfo{author}{\bibfnamefont{M.}~\bibnamefont{Goldhaber}} \bibnamefont{and}
  \bibinfo{author}{\bibfnamefont{E.}~\bibnamefont{Teller}},
  \bibinfo{journal}{Phys. Rev.} \textbf{\bibinfo{volume}{74}},
  \bibinfo{pages}{1046} (\bibinfo{year}{1948}),
  \urlprefix\url{https://link.aps.org/doi/10.1103/PhysRev.74.1046}.

\bibitem[{\citenamefont{Uhl and Kopecky}(1995)}]{Uhl94}
\bibinfo{author}{\bibfnamefont{M.}~\bibnamefont{Uhl}} \bibnamefont{and}
  \bibinfo{author}{\bibfnamefont{J.}~\bibnamefont{Kopecky}},
  \bibinfo{type}{Tech. Rep.} \bibinfo{number}{ECN-RX-94-099},
  \bibinfo{institution}{Netherlands} (\bibinfo{year}{1995}).

\bibitem[{\citenamefont{Kopecky and Uhl}(1990)}]{Kopecky90}
\bibinfo{author}{\bibfnamefont{J.}~\bibnamefont{Kopecky}} \bibnamefont{and}
  \bibinfo{author}{\bibfnamefont{M.}~\bibnamefont{Uhl}},
  \bibinfo{journal}{Phys. Rev. C} \textbf{\bibinfo{volume}{41}},
  \bibinfo{pages}{1941} (\bibinfo{year}{1990}),
  \urlprefix\url{https://link.aps.org/doi/10.1103/PhysRevC.41.1941}.

\bibitem[{\citenamefont{Brink}(1955)}]{Brink55}
\bibinfo{author}{\bibfnamefont{D.}~\bibnamefont{Brink}}, Ph.D. thesis,
  \bibinfo{school}{University of Oxford} (\bibinfo{year}{1955}).

\bibitem[{\citenamefont{Axel}(1962)}]{Axel62}
\bibinfo{author}{\bibfnamefont{P.}~\bibnamefont{Axel}}, \bibinfo{journal}{Phys.
  Rev.} \textbf{\bibinfo{volume}{126}}, \bibinfo{pages}{671}
  (\bibinfo{year}{1962}),
  \urlprefix\url{https://link.aps.org/doi/10.1103/PhysRev.126.671}.

\bibitem[{\citenamefont{Varlamov et~al.}(1999)\citenamefont{Varlamov, Varlamov,
  Rudenko, and Stepanov}}]{Varlamov99}
\bibinfo{author}{\bibfnamefont{A.}~\bibnamefont{Varlamov}},
  \bibinfo{author}{\bibfnamefont{V.}~\bibnamefont{Varlamov}},
  \bibinfo{author}{\bibfnamefont{D.}~\bibnamefont{Rudenko}}, \bibnamefont{and}
  \bibinfo{author}{\bibfnamefont{M.}~\bibnamefont{Stepanov}},
  \bibinfo{type}{Tech. Rep.}, \bibinfo{institution}{International Atomic Energy
  Agency. International Nuclear Data Committee} (\bibinfo{year}{1999}).

\bibitem[{\citenamefont{Kawano et~al.}(2020)\citenamefont{Kawano, Cho,
  Dimitriou, Filipescu, Iwamoto, Plujko, Tao, Utsunomiya, Varlamov, Xu
  et~al.}}]{Kawano19}
\bibinfo{author}{\bibfnamefont{T.}~\bibnamefont{Kawano}},
  \bibinfo{author}{\bibfnamefont{Y.}~\bibnamefont{Cho}},
  \bibinfo{author}{\bibfnamefont{P.}~\bibnamefont{Dimitriou}},
  \bibinfo{author}{\bibfnamefont{D.}~\bibnamefont{Filipescu}},
  \bibinfo{author}{\bibfnamefont{N.}~\bibnamefont{Iwamoto}},
  \bibinfo{author}{\bibfnamefont{V.}~\bibnamefont{Plujko}},
  \bibinfo{author}{\bibfnamefont{X.}~\bibnamefont{Tao}},
  \bibinfo{author}{\bibfnamefont{H.}~\bibnamefont{Utsunomiya}},
  \bibinfo{author}{\bibfnamefont{V.}~\bibnamefont{Varlamov}},
  \bibinfo{author}{\bibfnamefont{R.}~\bibnamefont{Xu}}, \bibnamefont{et~al.},
  \bibinfo{journal}{Nuclear Data Sheets} \textbf{\bibinfo{volume}{163}},
  \bibinfo{pages}{109 } (\bibinfo{year}{2020}), ISSN \bibinfo{issn}{0090-3752},
  \urlprefix\url{http://www.sciencedirect.com/science/article/pii/S0090375219300699}.

\bibitem[{\citenamefont{{Arruda Neto} and Herman}(1980)}]{Arruda80}
\bibinfo{author}{\bibfnamefont{J.}~\bibnamefont{{Arruda Neto}}}
  \bibnamefont{and} \bibinfo{author}{\bibfnamefont{B.}~\bibnamefont{Herman}},
  \bibinfo{journal}{Nuclear Physics A} \textbf{\bibinfo{volume}{349}},
  \bibinfo{pages}{483} (\bibinfo{year}{1980}), ISSN \bibinfo{issn}{0375-9474},
  \urlprefix\url{https://www.sciencedirect.com/science/article/pii/0375947480903024}.

\bibitem[{\citenamefont{Capote et~al.}(2009)\citenamefont{Capote, Herman,
  Obložinský, Young, Goriely, Belgya, Ignatyuk, Koning, Hilaire, Plujko
  et~al.}}]{Capote09}
\bibinfo{author}{\bibfnamefont{R.}~\bibnamefont{Capote}},
  \bibinfo{author}{\bibfnamefont{M.}~\bibnamefont{Herman}},
  \bibinfo{author}{\bibfnamefont{P.}~\bibnamefont{Obložinský}},
  \bibinfo{author}{\bibfnamefont{P.}~\bibnamefont{Young}},
  \bibinfo{author}{\bibfnamefont{S.}~\bibnamefont{Goriely}},
  \bibinfo{author}{\bibfnamefont{T.}~\bibnamefont{Belgya}},
  \bibinfo{author}{\bibfnamefont{A.}~\bibnamefont{Ignatyuk}},
  \bibinfo{author}{\bibfnamefont{A.}~\bibnamefont{Koning}},
  \bibinfo{author}{\bibfnamefont{S.}~\bibnamefont{Hilaire}},
  \bibinfo{author}{\bibfnamefont{V.}~\bibnamefont{Plujko}},
  \bibnamefont{et~al.}, \bibinfo{journal}{Nuclear Data Sheets}
  \textbf{\bibinfo{volume}{110}}, \bibinfo{pages}{3107} (\bibinfo{year}{2009}),
  ISSN \bibinfo{issn}{0090-3752}, \bibinfo{note}{special Issue on Nuclear
  Reaction Data},
  \urlprefix\url{https://www.sciencedirect.com/science/article/pii/S0090375209000994}.

\bibitem[{\citenamefont{Moszkowski}(1957)}]{Mosz57}
\bibinfo{author}{\bibfnamefont{S.}~\bibnamefont{Moszkowski}},
  \bibinfo{journal}{Handbuch der Physik} \textbf{\bibinfo{volume}{39}},
  \bibinfo{pages}{470} (\bibinfo{year}{1957}).

\bibitem[{\citenamefont{Blomqvist and Molinari}(1968)}]{Blomqvist68}
\bibinfo{author}{\bibfnamefont{J.}~\bibnamefont{Blomqvist}} \bibnamefont{and}
  \bibinfo{author}{\bibfnamefont{A.}~\bibnamefont{Molinari}},
  \bibinfo{journal}{Nuclear Physics A} \textbf{\bibinfo{volume}{106}},
  \bibinfo{pages}{545} (\bibinfo{year}{1968}), ISSN \bibinfo{issn}{0375-9474},
  \urlprefix\url{https://www.sciencedirect.com/science/article/pii/0375947468905150}.

\bibitem[{\citenamefont{Dodge}(2008)}]{COD}
\bibinfo{author}{\bibfnamefont{Y.}~\bibnamefont{Dodge}},
  \emph{\bibinfo{title}{Coefficient of Determination}}
  (\bibinfo{publisher}{Springer New York}, \bibinfo{address}{New York, NY},
  \bibinfo{year}{2008}), pp. \bibinfo{pages}{88--91}, ISBN
  \bibinfo{isbn}{978-0-387-32833-1},
  \urlprefix\url{https://doi.org/10.1007/978-0-387-32833-1_62}.

\bibitem[{\citenamefont{Faul et~al.}(1981)\citenamefont{Faul, Berman, Meyer,
  and Olson}}]{Faul81}
\bibinfo{author}{\bibfnamefont{D.~D.} \bibnamefont{Faul}},
  \bibinfo{author}{\bibfnamefont{B.~L.} \bibnamefont{Berman}},
  \bibinfo{author}{\bibfnamefont{P.}~\bibnamefont{Meyer}}, \bibnamefont{and}
  \bibinfo{author}{\bibfnamefont{D.~L.} \bibnamefont{Olson}},
  \bibinfo{journal}{Phys. Rev. C} \textbf{\bibinfo{volume}{24}},
  \bibinfo{pages}{849} (\bibinfo{year}{1981}),
  \urlprefix\url{https://link.aps.org/doi/10.1103/PhysRevC.24.849}.

\bibitem[{\citenamefont{Dytlewski et~al.}(1984)\citenamefont{Dytlewski,
  Siddiqui, and Thies}}]{Dytlewski84}
\bibinfo{author}{\bibfnamefont{N.}~\bibnamefont{Dytlewski}},
  \bibinfo{author}{\bibfnamefont{S.}~\bibnamefont{Siddiqui}}, \bibnamefont{and}
  \bibinfo{author}{\bibfnamefont{H.}~\bibnamefont{Thies}},
  \bibinfo{journal}{Nuclear Physics A} \textbf{\bibinfo{volume}{430}},
  \bibinfo{pages}{214} (\bibinfo{year}{1984}), ISSN \bibinfo{issn}{0375-9474},
  \urlprefix\url{https://www.sciencedirect.com/science/article/pii/037594748490201X}.

\bibitem[{\citenamefont{Pritychenko et~al.}(2017)\citenamefont{Pritychenko,
  Birch, and Singh}}]{Pritychenko17}
\bibinfo{author}{\bibfnamefont{B.}~\bibnamefont{Pritychenko}},
  \bibinfo{author}{\bibfnamefont{M.}~\bibnamefont{Birch}}, \bibnamefont{and}
  \bibinfo{author}{\bibfnamefont{B.}~\bibnamefont{Singh}},
  \bibinfo{journal}{Nuclear Physics A} \textbf{\bibinfo{volume}{962}},
  \bibinfo{pages}{73} (\bibinfo{year}{2017}), ISSN \bibinfo{issn}{0375-9474},
  \urlprefix\url{https://www.sciencedirect.com/science/article/pii/S037594741730057X}.

\bibitem[{\citenamefont{Möller et~al.}(2016)\citenamefont{Möller, Sierk,
  Ichikawa, and Sagawa}}]{Moller16}
\bibinfo{author}{\bibfnamefont{P.}~\bibnamefont{Möller}},
  \bibinfo{author}{\bibfnamefont{A.}~\bibnamefont{Sierk}},
  \bibinfo{author}{\bibfnamefont{T.}~\bibnamefont{Ichikawa}}, \bibnamefont{and}
  \bibinfo{author}{\bibfnamefont{H.}~\bibnamefont{Sagawa}},
  \bibinfo{journal}{Atomic Data and Nuclear Data Tables}
  \textbf{\bibinfo{volume}{109-110}}, \bibinfo{pages}{1–204}
  (\bibinfo{year}{2016}), ISSN \bibinfo{issn}{0092-640X},
  \urlprefix\url{http://dx.doi.org/10.1016/j.adt.2015.10.002}.

\bibitem[{\citenamefont{Firestone et~al.}(1996)\citenamefont{Firestone, Chu,
  Shirley, Baglin, and Zipkin}}]{Firestone96}
\bibinfo{author}{\bibfnamefont{R.}~\bibnamefont{Firestone}},
  \bibinfo{author}{\bibfnamefont{S.}~\bibnamefont{Chu}},
  \bibinfo{author}{\bibfnamefont{V.}~\bibnamefont{Shirley}},
  \bibinfo{author}{\bibfnamefont{C.}~\bibnamefont{Baglin}}, \bibnamefont{and}
  \bibinfo{author}{\bibfnamefont{J.}~\bibnamefont{Zipkin}},
  \emph{\bibinfo{title}{The 8th Edition of the {T}able of {I}sotopes}}
  (\bibinfo{publisher}{John Wiley and Sons}, \bibinfo{year}{1996}), ISBN
  \bibinfo{isbn}{0471356336}.

\bibitem[{\citenamefont{Li et~al.}(2021)\citenamefont{Li, Luo, and
  Wang}}]{Li21}
\bibinfo{author}{\bibfnamefont{T.}~\bibnamefont{Li}},
  \bibinfo{author}{\bibfnamefont{Y.}~\bibnamefont{Luo}}, \bibnamefont{and}
  \bibinfo{author}{\bibfnamefont{N.}~\bibnamefont{Wang}},
  \bibinfo{journal}{Atomic Data and Nuclear Data Tables}
  \textbf{\bibinfo{volume}{140}}, \bibinfo{pages}{101440}
  (\bibinfo{year}{2021}), ISSN \bibinfo{issn}{0092-640X},
  \urlprefix\url{https://www.sciencedirect.com/science/article/pii/S0092640X21000267}.

\bibitem[{\citenamefont{Flynn et~al.}(1974)\citenamefont{Flynn, Sherman, and
  Stein}}]{Flynn74}
\bibinfo{author}{\bibfnamefont{E.~R.} \bibnamefont{Flynn}},
  \bibinfo{author}{\bibfnamefont{J.}~\bibnamefont{Sherman}}, \bibnamefont{and}
  \bibinfo{author}{\bibfnamefont{N.}~\bibnamefont{Stein}},
  \bibinfo{journal}{Phys. Rev. Lett.} \textbf{\bibinfo{volume}{32}},
  \bibinfo{pages}{846} (\bibinfo{year}{1974}),
  \urlprefix\url{https://link.aps.org/doi/10.1103/PhysRevLett.32.846}.

\bibitem[{\citenamefont{Richter}(1983)}]{Richter83}
\bibinfo{author}{\bibfnamefont{A.}~\bibnamefont{Richter}},
  \bibinfo{journal}{Physica Scripta} \textbf{\bibinfo{volume}{T5}},
  \bibinfo{pages}{63} (\bibinfo{year}{1983}),
  \urlprefix\url{https://doi.org/10.1088%2F0031-8949%2F1983%2Ft5%2F010}.

\bibitem[{\citenamefont{Scheck et~al.}(2013)\citenamefont{Scheck, Ponomarev,
  Aumann, Beller, Fritzsche, Isaak, Kelley, Kwan, Pietralla, Raut
  et~al.}}]{Scheck13}
\bibinfo{author}{\bibfnamefont{M.}~\bibnamefont{Scheck}},
  \bibinfo{author}{\bibfnamefont{V.~Y.} \bibnamefont{Ponomarev}},
  \bibinfo{author}{\bibfnamefont{T.}~\bibnamefont{Aumann}},
  \bibinfo{author}{\bibfnamefont{J.}~\bibnamefont{Beller}},
  \bibinfo{author}{\bibfnamefont{M.}~\bibnamefont{Fritzsche}},
  \bibinfo{author}{\bibfnamefont{J.}~\bibnamefont{Isaak}},
  \bibinfo{author}{\bibfnamefont{J.~H.} \bibnamefont{Kelley}},
  \bibinfo{author}{\bibfnamefont{E.}~\bibnamefont{Kwan}},
  \bibinfo{author}{\bibfnamefont{N.}~\bibnamefont{Pietralla}},
  \bibinfo{author}{\bibfnamefont{R.}~\bibnamefont{Raut}}, \bibnamefont{et~al.},
  \bibinfo{journal}{Phys. Rev. C} \textbf{\bibinfo{volume}{87}},
  \bibinfo{pages}{051304} (\bibinfo{year}{2013}),
  \urlprefix\url{https://link.aps.org/doi/10.1103/PhysRevC.87.051304}.

\bibitem[{\citenamefont{Schramm et~al.}(2012)\citenamefont{Schramm, Massarczyk,
  Junghans, Belgya, Beyer, Birgersson, Grosse, Kempe, Kis, Kosev
  et~al.}}]{Schramm12}
\bibinfo{author}{\bibfnamefont{G.}~\bibnamefont{Schramm}},
  \bibinfo{author}{\bibfnamefont{R.}~\bibnamefont{Massarczyk}},
  \bibinfo{author}{\bibfnamefont{A.~R.} \bibnamefont{Junghans}},
  \bibinfo{author}{\bibfnamefont{T.}~\bibnamefont{Belgya}},
  \bibinfo{author}{\bibfnamefont{R.}~\bibnamefont{Beyer}},
  \bibinfo{author}{\bibfnamefont{E.}~\bibnamefont{Birgersson}},
  \bibinfo{author}{\bibfnamefont{E.}~\bibnamefont{Grosse}},
  \bibinfo{author}{\bibfnamefont{M.}~\bibnamefont{Kempe}},
  \bibinfo{author}{\bibfnamefont{Z.}~\bibnamefont{Kis}},
  \bibinfo{author}{\bibfnamefont{K.}~\bibnamefont{Kosev}},
  \bibnamefont{et~al.}, \bibinfo{journal}{Phys. Rev. C}
  \textbf{\bibinfo{volume}{85}}, \bibinfo{pages}{014311}
  (\bibinfo{year}{2012}),
  \urlprefix\url{https://link.aps.org/doi/10.1103/PhysRevC.85.014311}.

\bibitem[{\citenamefont{Schwengner et~al.}(2013)\citenamefont{Schwengner,
  Massarczyk, Rusev, Tsoneva, Bemmerer, Beyer, Hannaske, Junghans, Kelley, Kwan
  et~al.}}]{Schweng13}
\bibinfo{author}{\bibfnamefont{R.}~\bibnamefont{Schwengner}},
  \bibinfo{author}{\bibfnamefont{R.}~\bibnamefont{Massarczyk}},
  \bibinfo{author}{\bibfnamefont{G.}~\bibnamefont{Rusev}},
  \bibinfo{author}{\bibfnamefont{N.}~\bibnamefont{Tsoneva}},
  \bibinfo{author}{\bibfnamefont{D.}~\bibnamefont{Bemmerer}},
  \bibinfo{author}{\bibfnamefont{R.}~\bibnamefont{Beyer}},
  \bibinfo{author}{\bibfnamefont{R.}~\bibnamefont{Hannaske}},
  \bibinfo{author}{\bibfnamefont{A.~R.} \bibnamefont{Junghans}},
  \bibinfo{author}{\bibfnamefont{J.~H.} \bibnamefont{Kelley}},
  \bibinfo{author}{\bibfnamefont{E.}~\bibnamefont{Kwan}}, \bibnamefont{et~al.},
  \bibinfo{journal}{Phys. Rev. C} \textbf{\bibinfo{volume}{87}},
  \bibinfo{pages}{024306} (\bibinfo{year}{2013}),
  \urlprefix\url{https://link.aps.org/doi/10.1103/PhysRevC.87.024306}.

\bibitem[{\citenamefont{Schwengner et~al.}(2007)\citenamefont{Schwengner,
  Rusev, Benouaret, Beyer, Erhard, Grosse, Junghans, Klug, Kosev, Kostov
  et~al.}}]{Schweng07}
\bibinfo{author}{\bibfnamefont{R.}~\bibnamefont{Schwengner}},
  \bibinfo{author}{\bibfnamefont{G.}~\bibnamefont{Rusev}},
  \bibinfo{author}{\bibfnamefont{N.}~\bibnamefont{Benouaret}},
  \bibinfo{author}{\bibfnamefont{R.}~\bibnamefont{Beyer}},
  \bibinfo{author}{\bibfnamefont{M.}~\bibnamefont{Erhard}},
  \bibinfo{author}{\bibfnamefont{E.}~\bibnamefont{Grosse}},
  \bibinfo{author}{\bibfnamefont{A.~R.} \bibnamefont{Junghans}},
  \bibinfo{author}{\bibfnamefont{J.}~\bibnamefont{Klug}},
  \bibinfo{author}{\bibfnamefont{K.}~\bibnamefont{Kosev}},
  \bibinfo{author}{\bibfnamefont{L.}~\bibnamefont{Kostov}},
  \bibnamefont{et~al.}, \bibinfo{journal}{Phys. Rev. C}
  \textbf{\bibinfo{volume}{76}}, \bibinfo{pages}{034321}
  (\bibinfo{year}{2007}),
  \urlprefix\url{https://link.aps.org/doi/10.1103/PhysRevC.76.034321}.

\bibitem[{\citenamefont{Benouaret et~al.}(2009)\citenamefont{Benouaret,
  Schwengner, Rusev, D\"onau, Beyer, Erhard, Grosse, Junghans, Kosev, Nair
  et~al.}}]{Beno09}
\bibinfo{author}{\bibfnamefont{N.}~\bibnamefont{Benouaret}},
  \bibinfo{author}{\bibfnamefont{R.}~\bibnamefont{Schwengner}},
  \bibinfo{author}{\bibfnamefont{G.}~\bibnamefont{Rusev}},
  \bibinfo{author}{\bibfnamefont{F.}~\bibnamefont{D\"onau}},
  \bibinfo{author}{\bibfnamefont{R.}~\bibnamefont{Beyer}},
  \bibinfo{author}{\bibfnamefont{M.}~\bibnamefont{Erhard}},
  \bibinfo{author}{\bibfnamefont{E.}~\bibnamefont{Grosse}},
  \bibinfo{author}{\bibfnamefont{A.~R.} \bibnamefont{Junghans}},
  \bibinfo{author}{\bibfnamefont{K.}~\bibnamefont{Kosev}},
  \bibinfo{author}{\bibfnamefont{C.}~\bibnamefont{Nair}}, \bibnamefont{et~al.},
  \bibinfo{journal}{Phys. Rev. C} \textbf{\bibinfo{volume}{79}},
  \bibinfo{pages}{014303} (\bibinfo{year}{2009}),
  \urlprefix\url{https://link.aps.org/doi/10.1103/PhysRevC.79.014303}.

\bibitem[{\citenamefont{Kamerdzhiev and Kovalev}(2006)}]{Kamer06}
\bibinfo{author}{\bibfnamefont{S.}~\bibnamefont{Kamerdzhiev}} \bibnamefont{and}
  \bibinfo{author}{\bibfnamefont{S.}~\bibnamefont{Kovalev}},
  \bibinfo{journal}{Phyics of Atomic Nuclei} \textbf{\bibinfo{volume}{69}},
  \bibinfo{pages}{418} (\bibinfo{year}{2006}).

\bibitem[{\citenamefont{Iwamoto et~al.}(2012)\citenamefont{Iwamoto, Utsunomiya,
  Tamii, Akimune, Nakada, Shima, Yamagata, Kawabata, Fujita, Matsubara
  et~al.}}]{Iwam12}
\bibinfo{author}{\bibfnamefont{C.}~\bibnamefont{Iwamoto}},
  \bibinfo{author}{\bibfnamefont{H.}~\bibnamefont{Utsunomiya}},
  \bibinfo{author}{\bibfnamefont{A.}~\bibnamefont{Tamii}},
  \bibinfo{author}{\bibfnamefont{H.}~\bibnamefont{Akimune}},
  \bibinfo{author}{\bibfnamefont{H.}~\bibnamefont{Nakada}},
  \bibinfo{author}{\bibfnamefont{T.}~\bibnamefont{Shima}},
  \bibinfo{author}{\bibfnamefont{T.}~\bibnamefont{Yamagata}},
  \bibinfo{author}{\bibfnamefont{T.}~\bibnamefont{Kawabata}},
  \bibinfo{author}{\bibfnamefont{Y.}~\bibnamefont{Fujita}},
  \bibinfo{author}{\bibfnamefont{H.}~\bibnamefont{Matsubara}},
  \bibnamefont{et~al.}, \bibinfo{journal}{Phys. Rev. Lett.}
  \textbf{\bibinfo{volume}{108}}, \bibinfo{pages}{262501}
  (\bibinfo{year}{2012}),
  \urlprefix\url{https://link.aps.org/doi/10.1103/PhysRevLett.108.262501}.

\bibitem[{\citenamefont{Schwengner et~al.}(2008)\citenamefont{Schwengner,
  Rusev, Tsoneva, Benouaret, Beyer, Erhard, Grosse, Junghans, Klug, Kosev
  et~al.}}]{Schweng08}
\bibinfo{author}{\bibfnamefont{R.}~\bibnamefont{Schwengner}},
  \bibinfo{author}{\bibfnamefont{G.}~\bibnamefont{Rusev}},
  \bibinfo{author}{\bibfnamefont{N.}~\bibnamefont{Tsoneva}},
  \bibinfo{author}{\bibfnamefont{N.}~\bibnamefont{Benouaret}},
  \bibinfo{author}{\bibfnamefont{R.}~\bibnamefont{Beyer}},
  \bibinfo{author}{\bibfnamefont{M.}~\bibnamefont{Erhard}},
  \bibinfo{author}{\bibfnamefont{E.}~\bibnamefont{Grosse}},
  \bibinfo{author}{\bibfnamefont{A.~R.} \bibnamefont{Junghans}},
  \bibinfo{author}{\bibfnamefont{J.}~\bibnamefont{Klug}},
  \bibinfo{author}{\bibfnamefont{K.}~\bibnamefont{Kosev}},
  \bibnamefont{et~al.}, \bibinfo{journal}{Phys. Rev. C}
  \textbf{\bibinfo{volume}{78}}, \bibinfo{pages}{064314}
  (\bibinfo{year}{2008}),
  \urlprefix\url{https://link.aps.org/doi/10.1103/PhysRevC.78.064314}.

\bibitem[{\citenamefont{Derya et~al.}(2013)\citenamefont{Derya, Endres, Elvers,
  Harakeh, Pietralla, Romig, Savran, Scheck, Siebenhühner, Stoica
  et~al.}}]{Derya13}
\bibinfo{author}{\bibfnamefont{V.}~\bibnamefont{Derya}},
  \bibinfo{author}{\bibfnamefont{J.}~\bibnamefont{Endres}},
  \bibinfo{author}{\bibfnamefont{M.}~\bibnamefont{Elvers}},
  \bibinfo{author}{\bibfnamefont{M.}~\bibnamefont{Harakeh}},
  \bibinfo{author}{\bibfnamefont{N.}~\bibnamefont{Pietralla}},
  \bibinfo{author}{\bibfnamefont{C.}~\bibnamefont{Romig}},
  \bibinfo{author}{\bibfnamefont{D.}~\bibnamefont{Savran}},
  \bibinfo{author}{\bibfnamefont{M.}~\bibnamefont{Scheck}},
  \bibinfo{author}{\bibfnamefont{F.}~\bibnamefont{Siebenhühner}},
  \bibinfo{author}{\bibfnamefont{V.}~\bibnamefont{Stoica}},
  \bibnamefont{et~al.}, \bibinfo{journal}{Nuclear Physics A}
  \textbf{\bibinfo{volume}{906}}, \bibinfo{pages}{94 } (\bibinfo{year}{2013}),
  ISSN \bibinfo{issn}{0375-9474},
  \urlprefix\url{http://www.sciencedirect.com/science/article/pii/S0375947413001322}.

\bibitem[{\citenamefont{Eriksen et~al.}(2014)\citenamefont{Eriksen, Nyhus,
  Guttormsen, G\"orgen, Larsen, Renstr\o{}m, Ruud, Siem, Toft, Tveten
  et~al.}}]{Erik14}
\bibinfo{author}{\bibfnamefont{T.~K.} \bibnamefont{Eriksen}},
  \bibinfo{author}{\bibfnamefont{H.~T.} \bibnamefont{Nyhus}},
  \bibinfo{author}{\bibfnamefont{M.}~\bibnamefont{Guttormsen}},
  \bibinfo{author}{\bibfnamefont{A.}~\bibnamefont{G\"orgen}},
  \bibinfo{author}{\bibfnamefont{A.~C.} \bibnamefont{Larsen}},
  \bibinfo{author}{\bibfnamefont{T.}~\bibnamefont{Renstr\o{}m}},
  \bibinfo{author}{\bibfnamefont{I.~E.} \bibnamefont{Ruud}},
  \bibinfo{author}{\bibfnamefont{S.}~\bibnamefont{Siem}},
  \bibinfo{author}{\bibfnamefont{H.~K.} \bibnamefont{Toft}},
  \bibinfo{author}{\bibfnamefont{G.~M.} \bibnamefont{Tveten}},
  \bibnamefont{et~al.}, \bibinfo{journal}{Phys. Rev. C}
  \textbf{\bibinfo{volume}{90}}, \bibinfo{pages}{044311}
  (\bibinfo{year}{2014}),
  \urlprefix\url{https://link.aps.org/doi/10.1103/PhysRevC.90.044311}.

\bibitem[{\citenamefont{Toft et~al.}(2011)\citenamefont{Toft, Larsen, B\"urger,
  Guttormsen, G\"orgen, Nyhus, Renstr\o{}m, Siem, Tveten, and Voinov}}]{Toft11}
\bibinfo{author}{\bibfnamefont{H.~K.} \bibnamefont{Toft}},
  \bibinfo{author}{\bibfnamefont{A.~C.} \bibnamefont{Larsen}},
  \bibinfo{author}{\bibfnamefont{A.}~\bibnamefont{B\"urger}},
  \bibinfo{author}{\bibfnamefont{M.}~\bibnamefont{Guttormsen}},
  \bibinfo{author}{\bibfnamefont{A.}~\bibnamefont{G\"orgen}},
  \bibinfo{author}{\bibfnamefont{H.~T.} \bibnamefont{Nyhus}},
  \bibinfo{author}{\bibfnamefont{T.}~\bibnamefont{Renstr\o{}m}},
  \bibinfo{author}{\bibfnamefont{S.}~\bibnamefont{Siem}},
  \bibinfo{author}{\bibfnamefont{G.~M.} \bibnamefont{Tveten}},
  \bibnamefont{and} \bibinfo{author}{\bibfnamefont{A.}~\bibnamefont{Voinov}},
  \bibinfo{journal}{Phys. Rev. C} \textbf{\bibinfo{volume}{83}},
  \bibinfo{pages}{044320} (\bibinfo{year}{2011}),
  \urlprefix\url{https://link.aps.org/doi/10.1103/PhysRevC.83.044320}.

\bibitem[{\citenamefont{Endres et~al.}(2010)\citenamefont{Endres, Litvinova,
  Savran, Butler, Harakeh, Harissopulos, Herzberg, Kr\"ucken, Lagoyannis,
  Pietralla et~al.}}]{Endres10}
\bibinfo{author}{\bibfnamefont{J.}~\bibnamefont{Endres}},
  \bibinfo{author}{\bibfnamefont{E.}~\bibnamefont{Litvinova}},
  \bibinfo{author}{\bibfnamefont{D.}~\bibnamefont{Savran}},
  \bibinfo{author}{\bibfnamefont{P.~A.} \bibnamefont{Butler}},
  \bibinfo{author}{\bibfnamefont{M.~N.} \bibnamefont{Harakeh}},
  \bibinfo{author}{\bibfnamefont{S.}~\bibnamefont{Harissopulos}},
  \bibinfo{author}{\bibfnamefont{R.-D.} \bibnamefont{Herzberg}},
  \bibinfo{author}{\bibfnamefont{R.}~\bibnamefont{Kr\"ucken}},
  \bibinfo{author}{\bibfnamefont{A.}~\bibnamefont{Lagoyannis}},
  \bibinfo{author}{\bibfnamefont{N.}~\bibnamefont{Pietralla}},
  \bibnamefont{et~al.}, \bibinfo{journal}{Phys. Rev. Lett.}
  \textbf{\bibinfo{volume}{105}}, \bibinfo{pages}{212503}
  (\bibinfo{year}{2010}),
  \urlprefix\url{https://link.aps.org/doi/10.1103/PhysRevLett.105.212503}.

\bibitem[{\citenamefont{Massarczyk et~al.}(2014)\citenamefont{Massarczyk,
  Schwengner, D\"onau, Frauendorf, Anders, Bemmerer, Beyer, Bhatia, Birgersson,
  Butterling et~al.}}]{Mass14}
\bibinfo{author}{\bibfnamefont{R.}~\bibnamefont{Massarczyk}},
  \bibinfo{author}{\bibfnamefont{R.}~\bibnamefont{Schwengner}},
  \bibinfo{author}{\bibfnamefont{F.}~\bibnamefont{D\"onau}},
  \bibinfo{author}{\bibfnamefont{S.}~\bibnamefont{Frauendorf}},
  \bibinfo{author}{\bibfnamefont{M.}~\bibnamefont{Anders}},
  \bibinfo{author}{\bibfnamefont{D.}~\bibnamefont{Bemmerer}},
  \bibinfo{author}{\bibfnamefont{R.}~\bibnamefont{Beyer}},
  \bibinfo{author}{\bibfnamefont{C.}~\bibnamefont{Bhatia}},
  \bibinfo{author}{\bibfnamefont{E.}~\bibnamefont{Birgersson}},
  \bibinfo{author}{\bibfnamefont{M.}~\bibnamefont{Butterling}},
  \bibnamefont{et~al.}, \bibinfo{journal}{Phys. Rev. Lett.}
  \textbf{\bibinfo{volume}{112}}, \bibinfo{pages}{072501}
  (\bibinfo{year}{2014}),
  \urlprefix\url{https://link.aps.org/doi/10.1103/PhysRevLett.112.072501}.

\bibitem[{\citenamefont{Savran et~al.}(2008)\citenamefont{Savran, Fritzsche,
  Hasper, Lindenberg, M\"uller, Ponomarev, Sonnabend, and Zilges}}]{Savran08}
\bibinfo{author}{\bibfnamefont{D.}~\bibnamefont{Savran}},
  \bibinfo{author}{\bibfnamefont{M.}~\bibnamefont{Fritzsche}},
  \bibinfo{author}{\bibfnamefont{J.}~\bibnamefont{Hasper}},
  \bibinfo{author}{\bibfnamefont{K.}~\bibnamefont{Lindenberg}},
  \bibinfo{author}{\bibfnamefont{S.}~\bibnamefont{M\"uller}},
  \bibinfo{author}{\bibfnamefont{V.~Y.} \bibnamefont{Ponomarev}},
  \bibinfo{author}{\bibfnamefont{K.}~\bibnamefont{Sonnabend}},
  \bibnamefont{and} \bibinfo{author}{\bibfnamefont{A.}~\bibnamefont{Zilges}},
  \bibinfo{journal}{Phys. Rev. Lett.} \textbf{\bibinfo{volume}{100}},
  \bibinfo{pages}{232501} (\bibinfo{year}{2008}),
  \urlprefix\url{https://link.aps.org/doi/10.1103/PhysRevLett.100.232501}.

\bibitem[{\citenamefont{Endres et~al.}(2009)\citenamefont{Endres, Savran, Berg,
  Dendooven, Fritzsche, Harakeh, Hasper, W\"ortche, and Zilges}}]{Endres09}
\bibinfo{author}{\bibfnamefont{J.}~\bibnamefont{Endres}},
  \bibinfo{author}{\bibfnamefont{D.}~\bibnamefont{Savran}},
  \bibinfo{author}{\bibfnamefont{A.~M. v.~d.} \bibnamefont{Berg}},
  \bibinfo{author}{\bibfnamefont{P.}~\bibnamefont{Dendooven}},
  \bibinfo{author}{\bibfnamefont{M.}~\bibnamefont{Fritzsche}},
  \bibinfo{author}{\bibfnamefont{M.~N.} \bibnamefont{Harakeh}},
  \bibinfo{author}{\bibfnamefont{J.}~\bibnamefont{Hasper}},
  \bibinfo{author}{\bibfnamefont{H.~J.} \bibnamefont{W\"ortche}},
  \bibnamefont{and} \bibinfo{author}{\bibfnamefont{A.}~\bibnamefont{Zilges}},
  \bibinfo{journal}{Phys. Rev. C} \textbf{\bibinfo{volume}{80}},
  \bibinfo{pages}{034302} (\bibinfo{year}{2009}),
  \urlprefix\url{https://link.aps.org/doi/10.1103/PhysRevC.80.034302}.

\bibitem[{\citenamefont{Makinaga et~al.}(2010)\citenamefont{Makinaga,
  Schwengner, Rusev, D\"onau, Frauendorf, Bemmerer, Beyer, Crespo, Erhard,
  Junghans et~al.}}]{Makin10}
\bibinfo{author}{\bibfnamefont{A.}~\bibnamefont{Makinaga}},
  \bibinfo{author}{\bibfnamefont{R.}~\bibnamefont{Schwengner}},
  \bibinfo{author}{\bibfnamefont{G.}~\bibnamefont{Rusev}},
  \bibinfo{author}{\bibfnamefont{F.}~\bibnamefont{D\"onau}},
  \bibinfo{author}{\bibfnamefont{S.}~\bibnamefont{Frauendorf}},
  \bibinfo{author}{\bibfnamefont{D.}~\bibnamefont{Bemmerer}},
  \bibinfo{author}{\bibfnamefont{R.}~\bibnamefont{Beyer}},
  \bibinfo{author}{\bibfnamefont{P.}~\bibnamefont{Crespo}},
  \bibinfo{author}{\bibfnamefont{M.}~\bibnamefont{Erhard}},
  \bibinfo{author}{\bibfnamefont{A.~R.} \bibnamefont{Junghans}},
  \bibnamefont{et~al.}, \bibinfo{journal}{Phys. Rev. C}
  \textbf{\bibinfo{volume}{82}}, \bibinfo{pages}{024314}
  (\bibinfo{year}{2010}),
  \urlprefix\url{https://link.aps.org/doi/10.1103/PhysRevC.82.024314}.

\bibitem[{\citenamefont{Sarriguren et~al.}(1993)\citenamefont{Sarriguren,
  de~Guerra, Nojarov, and Faessler}}]{Sarri93}
\bibinfo{author}{\bibfnamefont{P.}~\bibnamefont{Sarriguren}},
  \bibinfo{author}{\bibfnamefont{E.~M.} \bibnamefont{de~Guerra}},
  \bibinfo{author}{\bibfnamefont{R.}~\bibnamefont{Nojarov}}, \bibnamefont{and}
  \bibinfo{author}{\bibfnamefont{A.}~\bibnamefont{Faessler}},
  \bibinfo{journal}{Journal of Physics G: Nuclear and Particle Physics}
  \textbf{\bibinfo{volume}{19}}, \bibinfo{pages}{291} (\bibinfo{year}{1993}),
  \urlprefix\url{https://doi.org/10.1088%2F0954-3899%2F19%2F2%2F011}.

\bibitem[{\citenamefont{Voinov et~al.}(2001)\citenamefont{Voinov, Guttormsen,
  Melby, Rekstad, Schiller, and Siem}}]{Voinov01}
\bibinfo{author}{\bibfnamefont{A.}~\bibnamefont{Voinov}},
  \bibinfo{author}{\bibfnamefont{M.}~\bibnamefont{Guttormsen}},
  \bibinfo{author}{\bibfnamefont{E.}~\bibnamefont{Melby}},
  \bibinfo{author}{\bibfnamefont{J.}~\bibnamefont{Rekstad}},
  \bibinfo{author}{\bibfnamefont{A.}~\bibnamefont{Schiller}}, \bibnamefont{and}
  \bibinfo{author}{\bibfnamefont{S.}~\bibnamefont{Siem}},
  \bibinfo{journal}{Phys. Rev. C} \textbf{\bibinfo{volume}{63}},
  \bibinfo{pages}{044313} (\bibinfo{year}{2001}),
  \urlprefix\url{https://link.aps.org/doi/10.1103/PhysRevC.63.044313}.

\bibitem[{\citenamefont{Lewis and Bertrand}(1972)}]{Lewis72}
\bibinfo{author}{\bibfnamefont{M.}~\bibnamefont{Lewis}} \bibnamefont{and}
  \bibinfo{author}{\bibfnamefont{F.}~\bibnamefont{Bertrand}},
  \bibinfo{journal}{Nuclear Physics A} \textbf{\bibinfo{volume}{196}},
  \bibinfo{pages}{337 } (\bibinfo{year}{1972}), ISSN \bibinfo{issn}{0375-9474},
  \urlprefix\url{http://www.sciencedirect.com/science/article/pii/0375947472909682}.

\bibitem[{\citenamefont{Poltoratska et~al.}(2012)\citenamefont{Poltoratska, von
  Neumann-Cosel, Tamii, Adachi, Bertulani, Carter, Dozono, Fujita, Fujita,
  Fujita et~al.}}]{Polto12}
\bibinfo{author}{\bibfnamefont{I.}~\bibnamefont{Poltoratska}},
  \bibinfo{author}{\bibfnamefont{P.}~\bibnamefont{von Neumann-Cosel}},
  \bibinfo{author}{\bibfnamefont{A.}~\bibnamefont{Tamii}},
  \bibinfo{author}{\bibfnamefont{T.}~\bibnamefont{Adachi}},
  \bibinfo{author}{\bibfnamefont{C.~A.} \bibnamefont{Bertulani}},
  \bibinfo{author}{\bibfnamefont{J.}~\bibnamefont{Carter}},
  \bibinfo{author}{\bibfnamefont{M.}~\bibnamefont{Dozono}},
  \bibinfo{author}{\bibfnamefont{H.}~\bibnamefont{Fujita}},
  \bibinfo{author}{\bibfnamefont{K.}~\bibnamefont{Fujita}},
  \bibinfo{author}{\bibfnamefont{Y.}~\bibnamefont{Fujita}},
  \bibnamefont{et~al.}, \bibinfo{journal}{Phys. Rev. C}
  \textbf{\bibinfo{volume}{85}}, \bibinfo{pages}{041304}
  (\bibinfo{year}{2012}),
  \urlprefix\url{https://link.aps.org/doi/10.1103/PhysRevC.85.041304}.

\bibitem[{\citenamefont{Laszewski et~al.}(1985)\citenamefont{Laszewski,
  Rullhusen, Hoblit, and LeBrun}}]{Lasze85}
\bibinfo{author}{\bibfnamefont{R.~M.} \bibnamefont{Laszewski}},
  \bibinfo{author}{\bibfnamefont{P.}~\bibnamefont{Rullhusen}},
  \bibinfo{author}{\bibfnamefont{S.~D.} \bibnamefont{Hoblit}},
  \bibnamefont{and} \bibinfo{author}{\bibfnamefont{S.~F.}
  \bibnamefont{LeBrun}}, \bibinfo{journal}{Phys. Rev. Lett.}
  \textbf{\bibinfo{volume}{54}}, \bibinfo{pages}{530} (\bibinfo{year}{1985}),
  \urlprefix\url{https://link.aps.org/doi/10.1103/PhysRevLett.54.530}.

\bibitem[{\citenamefont{Tornyi et~al.}(2014)\citenamefont{Tornyi, Guttormsen,
  Eriksen, G\"orgen, Giacoppo, Hagen, Krasznahorkay, Larsen, Renstr\o{}m, Rose
  et~al.}}]{Tornyi14}
\bibinfo{author}{\bibfnamefont{T.~G.} \bibnamefont{Tornyi}},
  \bibinfo{author}{\bibfnamefont{M.}~\bibnamefont{Guttormsen}},
  \bibinfo{author}{\bibfnamefont{T.~K.} \bibnamefont{Eriksen}},
  \bibinfo{author}{\bibfnamefont{A.}~\bibnamefont{G\"orgen}},
  \bibinfo{author}{\bibfnamefont{F.}~\bibnamefont{Giacoppo}},
  \bibinfo{author}{\bibfnamefont{T.~W.} \bibnamefont{Hagen}},
  \bibinfo{author}{\bibfnamefont{A.}~\bibnamefont{Krasznahorkay}},
  \bibinfo{author}{\bibfnamefont{A.~C.} \bibnamefont{Larsen}},
  \bibinfo{author}{\bibfnamefont{T.}~\bibnamefont{Renstr\o{}m}},
  \bibinfo{author}{\bibfnamefont{S.~J.} \bibnamefont{Rose}},
  \bibnamefont{et~al.}, \bibinfo{journal}{Phys. Rev. C}
  \textbf{\bibinfo{volume}{89}}, \bibinfo{pages}{044323}
  (\bibinfo{year}{2014}),
  \urlprefix\url{https://link.aps.org/doi/10.1103/PhysRevC.89.044323}.

\bibitem[{\citenamefont{Clark et~al.}(1999)\citenamefont{Clark, Lui,
  Youngblood, Bachtr, Garg, Harakeh, and Kalantar-Nayestanaki}}]{Clark99}
\bibinfo{author}{\bibfnamefont{H.}~\bibnamefont{Clark}},
  \bibinfo{author}{\bibfnamefont{Y.-W.} \bibnamefont{Lui}},
  \bibinfo{author}{\bibfnamefont{D.}~\bibnamefont{Youngblood}},
  \bibinfo{author}{\bibfnamefont{K.}~\bibnamefont{Bachtr}},
  \bibinfo{author}{\bibfnamefont{U.}~\bibnamefont{Garg}},
  \bibinfo{author}{\bibfnamefont{M.}~\bibnamefont{Harakeh}}, \bibnamefont{and}
  \bibinfo{author}{\bibfnamefont{N.}~\bibnamefont{Kalantar-Nayestanaki}},
  \bibinfo{journal}{Nuclear Physics A} \textbf{\bibinfo{volume}{649}},
  \bibinfo{pages}{57 } (\bibinfo{year}{1999}), ISSN \bibinfo{issn}{0375-9474},
  \bibinfo{note}{giant Resonances},
  \urlprefix\url{http://www.sciencedirect.com/science/article/pii/S0375947499000391}.

\bibitem[{\citenamefont{Bertrand et~al.}(1979)\citenamefont{Bertrand, Satchler,
  Horen, and [van~der Woude]}}]{Bert79}
\bibinfo{author}{\bibfnamefont{F.}~\bibnamefont{Bertrand}},
  \bibinfo{author}{\bibfnamefont{G.}~\bibnamefont{Satchler}},
  \bibinfo{author}{\bibfnamefont{D.}~\bibnamefont{Horen}}, \bibnamefont{and}
  \bibinfo{author}{\bibfnamefont{A.}~\bibnamefont{[van~der Woude]}},
  \bibinfo{journal}{Physics Letters B} \textbf{\bibinfo{volume}{80}},
  \bibinfo{pages}{198 } (\bibinfo{year}{1979}), ISSN \bibinfo{issn}{0370-2693},
  \urlprefix\url{http://www.sciencedirect.com/science/article/pii/0370269379901977}.

\bibitem[{\citenamefont{Lebrun et~al.}(1980)\citenamefont{Lebrun, Buenerd,
  Martin, [de Saintignon], and Perrin}}]{Lebrun80}
\bibinfo{author}{\bibfnamefont{D.}~\bibnamefont{Lebrun}},
  \bibinfo{author}{\bibfnamefont{M.}~\bibnamefont{Buenerd}},
  \bibinfo{author}{\bibfnamefont{P.}~\bibnamefont{Martin}},
  \bibinfo{author}{\bibfnamefont{P.}~\bibnamefont{[de Saintignon]}},
  \bibnamefont{and} \bibinfo{author}{\bibfnamefont{G.}~\bibnamefont{Perrin}},
  \bibinfo{journal}{Physics Letters B} \textbf{\bibinfo{volume}{97}},
  \bibinfo{pages}{358 } (\bibinfo{year}{1980}), ISSN \bibinfo{issn}{0370-2693},
  \urlprefix\url{http://www.sciencedirect.com/science/article/pii/037026938090619X}.

\bibitem[{\citenamefont{Patel et~al.}(2012)\citenamefont{Patel, Garg, Fujiwara,
  Akimune, Berg, Harakeh, Itoh, Kawabata, Kawase, Nayak et~al.}}]{Patel12}
\bibinfo{author}{\bibfnamefont{D.}~\bibnamefont{Patel}},
  \bibinfo{author}{\bibfnamefont{U.}~\bibnamefont{Garg}},
  \bibinfo{author}{\bibfnamefont{M.}~\bibnamefont{Fujiwara}},
  \bibinfo{author}{\bibfnamefont{H.}~\bibnamefont{Akimune}},
  \bibinfo{author}{\bibfnamefont{G.}~\bibnamefont{Berg}},
  \bibinfo{author}{\bibfnamefont{M.}~\bibnamefont{Harakeh}},
  \bibinfo{author}{\bibfnamefont{M.}~\bibnamefont{Itoh}},
  \bibinfo{author}{\bibfnamefont{T.}~\bibnamefont{Kawabata}},
  \bibinfo{author}{\bibfnamefont{K.}~\bibnamefont{Kawase}},
  \bibinfo{author}{\bibfnamefont{B.}~\bibnamefont{Nayak}},
  \bibnamefont{et~al.}, \bibinfo{journal}{Physics Letters B}
  \textbf{\bibinfo{volume}{718}}, \bibinfo{pages}{447 } (\bibinfo{year}{2012}),
  ISSN \bibinfo{issn}{0370-2693},
  \urlprefix\url{http://www.sciencedirect.com/science/article/pii/S0370269312011100}.

\bibitem[{\citenamefont{Youngblood
  et~al.}(1981{\natexlab{a}})\citenamefont{Youngblood, Bogucki, Bronson, Garg,
  Lui, and Rozsa}}]{Young97}
\bibinfo{author}{\bibfnamefont{D.~H.} \bibnamefont{Youngblood}},
  \bibinfo{author}{\bibfnamefont{P.}~\bibnamefont{Bogucki}},
  \bibinfo{author}{\bibfnamefont{J.~D.} \bibnamefont{Bronson}},
  \bibinfo{author}{\bibfnamefont{U.}~\bibnamefont{Garg}},
  \bibinfo{author}{\bibfnamefont{Y.~W.} \bibnamefont{Lui}}, \bibnamefont{and}
  \bibinfo{author}{\bibfnamefont{C.~M.} \bibnamefont{Rozsa}},
  \bibinfo{journal}{Phys. Rev. C} \textbf{\bibinfo{volume}{23}},
  \bibinfo{pages}{1997} (\bibinfo{year}{1981}{\natexlab{a}}),
  \urlprefix\url{https://link.aps.org/doi/10.1103/PhysRevC.23.1997}.

\bibitem[{\citenamefont{Lui et~al.}(1984)\citenamefont{Lui, Bogucki, Bronson,
  Youngblood, and Garg}}]{Lui84}
\bibinfo{author}{\bibfnamefont{Y.~W.} \bibnamefont{Lui}},
  \bibinfo{author}{\bibfnamefont{P.}~\bibnamefont{Bogucki}},
  \bibinfo{author}{\bibfnamefont{J.~D.} \bibnamefont{Bronson}},
  \bibinfo{author}{\bibfnamefont{D.~H.} \bibnamefont{Youngblood}},
  \bibnamefont{and} \bibinfo{author}{\bibfnamefont{U.}~\bibnamefont{Garg}},
  \bibinfo{journal}{Phys. Rev. C} \textbf{\bibinfo{volume}{30}},
  \bibinfo{pages}{51} (\bibinfo{year}{1984}),
  \urlprefix\url{https://link.aps.org/doi/10.1103/PhysRevC.30.51}.

\bibitem[{\citenamefont{Laszewski et~al.}(1986)\citenamefont{Laszewski,
  Rullhusen, Hoblit, and LeBrun}}]{Lasze13}
\bibinfo{author}{\bibfnamefont{R.~M.} \bibnamefont{Laszewski}},
  \bibinfo{author}{\bibfnamefont{P.}~\bibnamefont{Rullhusen}},
  \bibinfo{author}{\bibfnamefont{S.~D.} \bibnamefont{Hoblit}},
  \bibnamefont{and} \bibinfo{author}{\bibfnamefont{S.~F.}
  \bibnamefont{LeBrun}}, \bibinfo{journal}{Phys. Rev. C}
  \textbf{\bibinfo{volume}{34}}, \bibinfo{pages}{2013} (\bibinfo{year}{1986}),
  \urlprefix\url{https://link.aps.org/doi/10.1103/PhysRevC.34.2013}.

\bibitem[{\citenamefont{Sharma et~al.}(1988)\citenamefont{Sharma, Borghols,
  Brandenburg, Crona, van~der Woude, and Harakeh}}]{Sharma88}
\bibinfo{author}{\bibfnamefont{M.~M.} \bibnamefont{Sharma}},
  \bibinfo{author}{\bibfnamefont{W.~T.~A.} \bibnamefont{Borghols}},
  \bibinfo{author}{\bibfnamefont{S.}~\bibnamefont{Brandenburg}},
  \bibinfo{author}{\bibfnamefont{S.}~\bibnamefont{Crona}},
  \bibinfo{author}{\bibfnamefont{A.}~\bibnamefont{van~der Woude}},
  \bibnamefont{and} \bibinfo{author}{\bibfnamefont{M.~N.}
  \bibnamefont{Harakeh}}, \bibinfo{journal}{Phys. Rev. C}
  \textbf{\bibinfo{volume}{38}}, \bibinfo{pages}{2562} (\bibinfo{year}{1988}),
  \urlprefix\url{https://link.aps.org/doi/10.1103/PhysRevC.38.2562}.

\bibitem[{\citenamefont{Youngblood et~al.}(2002)\citenamefont{Youngblood, Lui,
  and Clark}}]{Young02}
\bibinfo{author}{\bibfnamefont{D.~H.} \bibnamefont{Youngblood}},
  \bibinfo{author}{\bibfnamefont{Y.-W.} \bibnamefont{Lui}}, \bibnamefont{and}
  \bibinfo{author}{\bibfnamefont{H.~L.} \bibnamefont{Clark}},
  \bibinfo{journal}{Phys. Rev. C} \textbf{\bibinfo{volume}{65}},
  \bibinfo{pages}{034302} (\bibinfo{year}{2002}),
  \urlprefix\url{https://link.aps.org/doi/10.1103/PhysRevC.65.034302}.

\bibitem[{\citenamefont{Itoh et~al.}(2003{\natexlab{a}})\citenamefont{Itoh,
  Sakaguchi, Uchida, Ishikawa, Kawabata, Murakami, Takeda, Taki, Terashima,
  Tsukahara et~al.}}]{Itoh03}
\bibinfo{author}{\bibfnamefont{M.}~\bibnamefont{Itoh}},
  \bibinfo{author}{\bibfnamefont{H.}~\bibnamefont{Sakaguchi}},
  \bibinfo{author}{\bibfnamefont{M.}~\bibnamefont{Uchida}},
  \bibinfo{author}{\bibfnamefont{T.}~\bibnamefont{Ishikawa}},
  \bibinfo{author}{\bibfnamefont{T.}~\bibnamefont{Kawabata}},
  \bibinfo{author}{\bibfnamefont{T.}~\bibnamefont{Murakami}},
  \bibinfo{author}{\bibfnamefont{H.}~\bibnamefont{Takeda}},
  \bibinfo{author}{\bibfnamefont{T.}~\bibnamefont{Taki}},
  \bibinfo{author}{\bibfnamefont{S.}~\bibnamefont{Terashima}},
  \bibinfo{author}{\bibfnamefont{N.}~\bibnamefont{Tsukahara}},
  \bibnamefont{et~al.}, \bibinfo{journal}{Phys. Rev. C}
  \textbf{\bibinfo{volume}{68}}, \bibinfo{pages}{064602}
  (\bibinfo{year}{2003}{\natexlab{a}}),
  \urlprefix\url{https://link.aps.org/doi/10.1103/PhysRevC.68.064602}.

\bibitem[{\citenamefont{Youngblood et~al.}(2004)\citenamefont{Youngblood, Lui,
  Clark, John, Tokimoto, and Chen}}]{Young04}
\bibinfo{author}{\bibfnamefont{D.~H.} \bibnamefont{Youngblood}},
  \bibinfo{author}{\bibfnamefont{Y.-W.} \bibnamefont{Lui}},
  \bibinfo{author}{\bibfnamefont{H.~L.} \bibnamefont{Clark}},
  \bibinfo{author}{\bibfnamefont{B.}~\bibnamefont{John}},
  \bibinfo{author}{\bibfnamefont{Y.}~\bibnamefont{Tokimoto}}, \bibnamefont{and}
  \bibinfo{author}{\bibfnamefont{X.}~\bibnamefont{Chen}},
  \bibinfo{journal}{Phys. Rev. C} \textbf{\bibinfo{volume}{69}},
  \bibinfo{pages}{034315} (\bibinfo{year}{2004}),
  \urlprefix\url{https://link.aps.org/doi/10.1103/PhysRevC.69.034315}.

\bibitem[{\citenamefont{Lui et~al.}(2004)\citenamefont{Lui, Youngblood,
  Tokimoto, Clark, and John}}]{Lui04}
\bibinfo{author}{\bibfnamefont{Y.-W.} \bibnamefont{Lui}},
  \bibinfo{author}{\bibfnamefont{D.~H.} \bibnamefont{Youngblood}},
  \bibinfo{author}{\bibfnamefont{Y.}~\bibnamefont{Tokimoto}},
  \bibinfo{author}{\bibfnamefont{H.~L.} \bibnamefont{Clark}}, \bibnamefont{and}
  \bibinfo{author}{\bibfnamefont{B.}~\bibnamefont{John}},
  \bibinfo{journal}{Phys. Rev. C} \textbf{\bibinfo{volume}{70}},
  \bibinfo{pages}{014307} (\bibinfo{year}{2004}),
  \urlprefix\url{https://link.aps.org/doi/10.1103/PhysRevC.70.014307}.

\bibitem[{\citenamefont{Lui et~al.}(2006)\citenamefont{Lui, Youngblood, Clark,
  Tokimoto, and John}}]{Lui06}
\bibinfo{author}{\bibfnamefont{Y.-W.} \bibnamefont{Lui}},
  \bibinfo{author}{\bibfnamefont{D.~H.} \bibnamefont{Youngblood}},
  \bibinfo{author}{\bibfnamefont{H.~L.} \bibnamefont{Clark}},
  \bibinfo{author}{\bibfnamefont{Y.}~\bibnamefont{Tokimoto}}, \bibnamefont{and}
  \bibinfo{author}{\bibfnamefont{B.}~\bibnamefont{John}},
  \bibinfo{journal}{Phys. Rev. C} \textbf{\bibinfo{volume}{73}},
  \bibinfo{pages}{014314} (\bibinfo{year}{2006}),
  \urlprefix\url{https://link.aps.org/doi/10.1103/PhysRevC.73.014314}.

\bibitem[{\citenamefont{Monrozeau et~al.}(2008)\citenamefont{Monrozeau, Khan,
  Blumenfeld, Demonchy, Mittig, Roussel-Chomaz, Beaumel, Caama\~no,
  Cortina-Gil, Ebran et~al.}}]{Monro08}
\bibinfo{author}{\bibfnamefont{C.}~\bibnamefont{Monrozeau}},
  \bibinfo{author}{\bibfnamefont{E.}~\bibnamefont{Khan}},
  \bibinfo{author}{\bibfnamefont{Y.}~\bibnamefont{Blumenfeld}},
  \bibinfo{author}{\bibfnamefont{C.~E.} \bibnamefont{Demonchy}},
  \bibinfo{author}{\bibfnamefont{W.}~\bibnamefont{Mittig}},
  \bibinfo{author}{\bibfnamefont{P.}~\bibnamefont{Roussel-Chomaz}},
  \bibinfo{author}{\bibfnamefont{D.}~\bibnamefont{Beaumel}},
  \bibinfo{author}{\bibfnamefont{M.}~\bibnamefont{Caama\~no}},
  \bibinfo{author}{\bibfnamefont{D.}~\bibnamefont{Cortina-Gil}},
  \bibinfo{author}{\bibfnamefont{J.~P.} \bibnamefont{Ebran}},
  \bibnamefont{et~al.}, \bibinfo{journal}{Phys. Rev. Lett.}
  \textbf{\bibinfo{volume}{100}}, \bibinfo{pages}{042501}
  (\bibinfo{year}{2008}),
  \urlprefix\url{https://link.aps.org/doi/10.1103/PhysRevLett.100.042501}.

\bibitem[{\citenamefont{Doering et~al.}(1975)\citenamefont{Doering, Galonsky,
  Patterson, and Bertsch}}]{Doer76}
\bibinfo{author}{\bibfnamefont{R.~R.} \bibnamefont{Doering}},
  \bibinfo{author}{\bibfnamefont{A.}~\bibnamefont{Galonsky}},
  \bibinfo{author}{\bibfnamefont{D.~M.} \bibnamefont{Patterson}},
  \bibnamefont{and} \bibinfo{author}{\bibfnamefont{G.~F.}
  \bibnamefont{Bertsch}}, \bibinfo{journal}{Phys. Rev. Lett.}
  \textbf{\bibinfo{volume}{35}}, \bibinfo{pages}{1691} (\bibinfo{year}{1975}),
  \urlprefix\url{https://link.aps.org/doi/10.1103/PhysRevLett.35.1691}.

\bibitem[{\citenamefont{Clark et~al.}(2001)\citenamefont{Clark, Lui, and
  Youngblood}}]{Clark01}
\bibinfo{author}{\bibfnamefont{H.~L.} \bibnamefont{Clark}},
  \bibinfo{author}{\bibfnamefont{Y.-W.} \bibnamefont{Lui}}, \bibnamefont{and}
  \bibinfo{author}{\bibfnamefont{D.~H.} \bibnamefont{Youngblood}},
  \bibinfo{journal}{Phys. Rev. C} \textbf{\bibinfo{volume}{63}},
  \bibinfo{pages}{031301} (\bibinfo{year}{2001}),
  \urlprefix\url{https://link.aps.org/doi/10.1103/PhysRevC.63.031301}.

\bibitem[{\citenamefont{Uchida et~al.}(2004)\citenamefont{Uchida, Sakaguchi,
  Itoh, Yosoi, Kawabata, Yasuda, Takeda, Murakami, Terashima, Kishi
  et~al.}}]{Uchi04}
\bibinfo{author}{\bibfnamefont{M.}~\bibnamefont{Uchida}},
  \bibinfo{author}{\bibfnamefont{H.}~\bibnamefont{Sakaguchi}},
  \bibinfo{author}{\bibfnamefont{M.}~\bibnamefont{Itoh}},
  \bibinfo{author}{\bibfnamefont{M.}~\bibnamefont{Yosoi}},
  \bibinfo{author}{\bibfnamefont{T.}~\bibnamefont{Kawabata}},
  \bibinfo{author}{\bibfnamefont{Y.}~\bibnamefont{Yasuda}},
  \bibinfo{author}{\bibfnamefont{H.}~\bibnamefont{Takeda}},
  \bibinfo{author}{\bibfnamefont{T.}~\bibnamefont{Murakami}},
  \bibinfo{author}{\bibfnamefont{S.}~\bibnamefont{Terashima}},
  \bibinfo{author}{\bibfnamefont{S.}~\bibnamefont{Kishi}},
  \bibnamefont{et~al.}, \bibinfo{journal}{Phys. Rev. C}
  \textbf{\bibinfo{volume}{69}}, \bibinfo{pages}{051301}
  (\bibinfo{year}{2004}),
  \urlprefix\url{https://link.aps.org/doi/10.1103/PhysRevC.69.051301}.

\bibitem[{\citenamefont{Li et~al.}(2010)\citenamefont{Li, Garg, Liu, Marks,
  Nayak, Madhusudhana~Rao, Fujiwara, Hashimoto, Nakanishi, Okumura
  et~al.}}]{Li10}
\bibinfo{author}{\bibfnamefont{T.}~\bibnamefont{Li}},
  \bibinfo{author}{\bibfnamefont{U.}~\bibnamefont{Garg}},
  \bibinfo{author}{\bibfnamefont{Y.}~\bibnamefont{Liu}},
  \bibinfo{author}{\bibfnamefont{R.}~\bibnamefont{Marks}},
  \bibinfo{author}{\bibfnamefont{B.~K.} \bibnamefont{Nayak}},
  \bibinfo{author}{\bibfnamefont{P.~V.} \bibnamefont{Madhusudhana~Rao}},
  \bibinfo{author}{\bibfnamefont{M.}~\bibnamefont{Fujiwara}},
  \bibinfo{author}{\bibfnamefont{H.}~\bibnamefont{Hashimoto}},
  \bibinfo{author}{\bibfnamefont{K.}~\bibnamefont{Nakanishi}},
  \bibinfo{author}{\bibfnamefont{S.}~\bibnamefont{Okumura}},
  \bibnamefont{et~al.}, \bibinfo{journal}{Phys. Rev. C}
  \textbf{\bibinfo{volume}{81}}, \bibinfo{pages}{034309}
  (\bibinfo{year}{2010}),
  \urlprefix\url{https://link.aps.org/doi/10.1103/PhysRevC.81.034309}.

\bibitem[{\citenamefont{Gupta et~al.}(2018)\citenamefont{Gupta, Howard, Garg,
  Matta, \ifmmode \mbox{\c{S}}\else \c{S}\fi{}enyi\ifmmode~\breve{g}\else
  \u{g}\fi{}it, Itoh, Ando, Aoki, Uchiyama, Adachi et~al.}}]{Gupta18}
\bibinfo{author}{\bibfnamefont{Y.~K.} \bibnamefont{Gupta}},
  \bibinfo{author}{\bibfnamefont{K.~B.} \bibnamefont{Howard}},
  \bibinfo{author}{\bibfnamefont{U.}~\bibnamefont{Garg}},
  \bibinfo{author}{\bibfnamefont{J.~T.} \bibnamefont{Matta}},
  \bibinfo{author}{\bibfnamefont{M.}~\bibnamefont{\ifmmode \mbox{\c{S}}\else
  \c{S}\fi{}enyi\ifmmode~\breve{g}\else \u{g}\fi{}it}},
  \bibinfo{author}{\bibfnamefont{M.}~\bibnamefont{Itoh}},
  \bibinfo{author}{\bibfnamefont{S.}~\bibnamefont{Ando}},
  \bibinfo{author}{\bibfnamefont{T.}~\bibnamefont{Aoki}},
  \bibinfo{author}{\bibfnamefont{A.}~\bibnamefont{Uchiyama}},
  \bibinfo{author}{\bibfnamefont{S.}~\bibnamefont{Adachi}},
  \bibnamefont{et~al.}, \bibinfo{journal}{Phys. Rev. C}
  \textbf{\bibinfo{volume}{97}}, \bibinfo{pages}{064323}
  (\bibinfo{year}{2018}),
  \urlprefix\url{https://link.aps.org/doi/10.1103/PhysRevC.97.064323}.

\bibitem[{\citenamefont{Schmidt et~al.}(1993)\citenamefont{Schmidt, Blaich,
  Elze, Emling, Freiesleben, Grimm, Henning, Holzmann, Keller, Klingler
  et~al.}}]{Schmidt93}
\bibinfo{author}{\bibfnamefont{R.}~\bibnamefont{Schmidt}},
  \bibinfo{author}{\bibfnamefont{T.}~\bibnamefont{Blaich}},
  \bibinfo{author}{\bibfnamefont{T.~W.} \bibnamefont{Elze}},
  \bibinfo{author}{\bibfnamefont{H.}~\bibnamefont{Emling}},
  \bibinfo{author}{\bibfnamefont{H.}~\bibnamefont{Freiesleben}},
  \bibinfo{author}{\bibfnamefont{K.}~\bibnamefont{Grimm}},
  \bibinfo{author}{\bibfnamefont{W.}~\bibnamefont{Henning}},
  \bibinfo{author}{\bibfnamefont{R.}~\bibnamefont{Holzmann}},
  \bibinfo{author}{\bibfnamefont{J.~G.} \bibnamefont{Keller}},
  \bibinfo{author}{\bibfnamefont{H.}~\bibnamefont{Klingler}},
  \bibnamefont{et~al.}, \bibinfo{journal}{Phys. Rev. Lett.}
  \textbf{\bibinfo{volume}{70}}, \bibinfo{pages}{1767} (\bibinfo{year}{1993}),
  \urlprefix\url{https://link.aps.org/doi/10.1103/PhysRevLett.70.1767}.

\bibitem[{\citenamefont{Davis et~al.}(1997)\citenamefont{Davis, Garg, Reviol,
  Harakeh, Bacher, Berg, Foster, Stephenson, Wang, J\"anecke et~al.}}]{Davis97}
\bibinfo{author}{\bibfnamefont{B.~F.} \bibnamefont{Davis}},
  \bibinfo{author}{\bibfnamefont{U.}~\bibnamefont{Garg}},
  \bibinfo{author}{\bibfnamefont{W.}~\bibnamefont{Reviol}},
  \bibinfo{author}{\bibfnamefont{M.~N.} \bibnamefont{Harakeh}},
  \bibinfo{author}{\bibfnamefont{A.}~\bibnamefont{Bacher}},
  \bibinfo{author}{\bibfnamefont{G.~P.~A.} \bibnamefont{Berg}},
  \bibinfo{author}{\bibfnamefont{C.~C.} \bibnamefont{Foster}},
  \bibinfo{author}{\bibfnamefont{E.~J.} \bibnamefont{Stephenson}},
  \bibinfo{author}{\bibfnamefont{Y.}~\bibnamefont{Wang}},
  \bibinfo{author}{\bibfnamefont{J.}~\bibnamefont{J\"anecke}},
  \bibnamefont{et~al.}, \bibinfo{journal}{Phys. Rev. Lett.}
  \textbf{\bibinfo{volume}{79}}, \bibinfo{pages}{609} (\bibinfo{year}{1997}),
  \urlprefix\url{https://link.aps.org/doi/10.1103/PhysRevLett.79.609}.

\bibitem[{\citenamefont{Youngblood
  et~al.}(1981{\natexlab{b}})\citenamefont{Youngblood, Bogucki, Bronson, Garg,
  Lui, and Rozsa}}]{Young81}
\bibinfo{author}{\bibfnamefont{D.~H.} \bibnamefont{Youngblood}},
  \bibinfo{author}{\bibfnamefont{P.}~\bibnamefont{Bogucki}},
  \bibinfo{author}{\bibfnamefont{J.~D.} \bibnamefont{Bronson}},
  \bibinfo{author}{\bibfnamefont{U.}~\bibnamefont{Garg}},
  \bibinfo{author}{\bibfnamefont{Y.~W.} \bibnamefont{Lui}}, \bibnamefont{and}
  \bibinfo{author}{\bibfnamefont{C.~M.} \bibnamefont{Rozsa}},
  \bibinfo{journal}{Phys. Rev. C} \textbf{\bibinfo{volume}{23}},
  \bibinfo{pages}{1997} (\bibinfo{year}{1981}{\natexlab{b}}),
  \urlprefix\url{https://link.aps.org/doi/10.1103/PhysRevC.23.1997}.

\bibitem[{\citenamefont{Sziklai et~al.}(1984)\citenamefont{Sziklai, Cameron,
  and Sz\"oghy}}]{Szik84}
\bibinfo{author}{\bibfnamefont{J.}~\bibnamefont{Sziklai}},
  \bibinfo{author}{\bibfnamefont{J.~A.} \bibnamefont{Cameron}},
  \bibnamefont{and} \bibinfo{author}{\bibfnamefont{I.~M.}
  \bibnamefont{Sz\"oghy}}, \bibinfo{journal}{Phys. Rev. C}
  \textbf{\bibinfo{volume}{30}}, \bibinfo{pages}{490} (\bibinfo{year}{1984}),
  \urlprefix\url{https://link.aps.org/doi/10.1103/PhysRevC.30.490}.

\bibitem[{\citenamefont{Morsch et~al.}(1982)\citenamefont{Morsch, Rogge, Turek,
  Decowski, Zemło, Mayer-Böricke, Martin, Berg, Katayama, Meissburger
  et~al.}}]{Morsch82}
\bibinfo{author}{\bibfnamefont{H.}~\bibnamefont{Morsch}},
  \bibinfo{author}{\bibfnamefont{M.}~\bibnamefont{Rogge}},
  \bibinfo{author}{\bibfnamefont{P.}~\bibnamefont{Turek}},
  \bibinfo{author}{\bibfnamefont{P.}~\bibnamefont{Decowski}},
  \bibinfo{author}{\bibfnamefont{L.}~\bibnamefont{Zemło}},
  \bibinfo{author}{\bibfnamefont{C.}~\bibnamefont{Mayer-Böricke}},
  \bibinfo{author}{\bibfnamefont{S.}~\bibnamefont{Martin}},
  \bibinfo{author}{\bibfnamefont{G.}~\bibnamefont{Berg}},
  \bibinfo{author}{\bibfnamefont{I.}~\bibnamefont{Katayama}},
  \bibinfo{author}{\bibfnamefont{J.}~\bibnamefont{Meissburger}},
  \bibnamefont{et~al.}, \bibinfo{journal}{Physics Letters B}
  \textbf{\bibinfo{volume}{119}}, \bibinfo{pages}{311 } (\bibinfo{year}{1982}),
  ISSN \bibinfo{issn}{0370-2693},
  \urlprefix\url{http://www.sciencedirect.com/science/article/pii/0370269382906773}.

\bibitem[{\citenamefont{Yamagata et~al.}(1981)\citenamefont{Yamagata,
  Kishimoto, Yuasa, Iwamoto, Saeki, Tanaka, Fukuda, Miura, Inoue, and
  Ogata}}]{Yama81}
\bibinfo{author}{\bibfnamefont{T.}~\bibnamefont{Yamagata}},
  \bibinfo{author}{\bibfnamefont{S.}~\bibnamefont{Kishimoto}},
  \bibinfo{author}{\bibfnamefont{K.}~\bibnamefont{Yuasa}},
  \bibinfo{author}{\bibfnamefont{K.}~\bibnamefont{Iwamoto}},
  \bibinfo{author}{\bibfnamefont{B.}~\bibnamefont{Saeki}},
  \bibinfo{author}{\bibfnamefont{M.}~\bibnamefont{Tanaka}},
  \bibinfo{author}{\bibfnamefont{T.}~\bibnamefont{Fukuda}},
  \bibinfo{author}{\bibfnamefont{I.}~\bibnamefont{Miura}},
  \bibinfo{author}{\bibfnamefont{M.}~\bibnamefont{Inoue}}, \bibnamefont{and}
  \bibinfo{author}{\bibfnamefont{H.}~\bibnamefont{Ogata}},
  \bibinfo{journal}{Phys. Rev. C} \textbf{\bibinfo{volume}{23}},
  \bibinfo{pages}{937} (\bibinfo{year}{1981}),
  \urlprefix\url{https://link.aps.org/doi/10.1103/PhysRevC.23.937}.

\bibitem[{\citenamefont{Saito et~al.}(1983)\citenamefont{Saito, Fujii, Saito,
  Torizuka, Tohei, and Hirota}}]{Saito83}
\bibinfo{author}{\bibfnamefont{T.}~\bibnamefont{Saito}},
  \bibinfo{author}{\bibfnamefont{Y.}~\bibnamefont{Fujii}},
  \bibinfo{author}{\bibfnamefont{K.}~\bibnamefont{Saito}},
  \bibinfo{author}{\bibfnamefont{Y.}~\bibnamefont{Torizuka}},
  \bibinfo{author}{\bibfnamefont{T.}~\bibnamefont{Tohei}}, \bibnamefont{and}
  \bibinfo{author}{\bibfnamefont{J.}~\bibnamefont{Hirota}},
  \bibinfo{journal}{Phys. Rev. C} \textbf{\bibinfo{volume}{28}},
  \bibinfo{pages}{652} (\bibinfo{year}{1983}),
  \urlprefix\url{https://link.aps.org/doi/10.1103/PhysRevC.28.652}.

\bibitem[{\citenamefont{Pitthan et~al.}(1980)\citenamefont{Pitthan, Buskirk,
  Houk, and Moore}}]{Pitt80}
\bibinfo{author}{\bibfnamefont{R.}~\bibnamefont{Pitthan}},
  \bibinfo{author}{\bibfnamefont{F.~R.} \bibnamefont{Buskirk}},
  \bibinfo{author}{\bibfnamefont{W.~A.} \bibnamefont{Houk}}, \bibnamefont{and}
  \bibinfo{author}{\bibfnamefont{R.~W.} \bibnamefont{Moore}},
  \bibinfo{journal}{Phys. Rev. C} \textbf{\bibinfo{volume}{21}},
  \bibinfo{pages}{28} (\bibinfo{year}{1980}),
  \urlprefix\url{https://link.aps.org/doi/10.1103/PhysRevC.21.28}.

\bibitem[{\citenamefont{Carey et~al.}(1980)\citenamefont{Carey, Cornelius,
  DiGiacomo, Moss, Adams, McClelland, Pauletta, Whitten, Gazzaly, Hintz
  et~al.}}]{Carey80}
\bibinfo{author}{\bibfnamefont{T.~A.} \bibnamefont{Carey}},
  \bibinfo{author}{\bibfnamefont{W.~D.} \bibnamefont{Cornelius}},
  \bibinfo{author}{\bibfnamefont{N.~J.} \bibnamefont{DiGiacomo}},
  \bibinfo{author}{\bibfnamefont{J.~M.} \bibnamefont{Moss}},
  \bibinfo{author}{\bibfnamefont{G.~S.} \bibnamefont{Adams}},
  \bibinfo{author}{\bibfnamefont{J.~B.} \bibnamefont{McClelland}},
  \bibinfo{author}{\bibfnamefont{G.}~\bibnamefont{Pauletta}},
  \bibinfo{author}{\bibfnamefont{C.}~\bibnamefont{Whitten}},
  \bibinfo{author}{\bibfnamefont{M.}~\bibnamefont{Gazzaly}},
  \bibinfo{author}{\bibfnamefont{N.}~\bibnamefont{Hintz}},
  \bibnamefont{et~al.}, \bibinfo{journal}{Phys. Rev. Lett.}
  \textbf{\bibinfo{volume}{45}}, \bibinfo{pages}{239} (\bibinfo{year}{1980}),
  \urlprefix\url{https://link.aps.org/doi/10.1103/PhysRevLett.45.239}.

\bibitem[{\citenamefont{Itoh et~al.}(2004)\citenamefont{Itoh, Sakaguchi,
  Uchida, Ishikawa, Kawabata, Murakami, Takeda, Taki, Terashima, Tsukahara
  et~al.}}]{Itoh04}
\bibinfo{author}{\bibfnamefont{M.}~\bibnamefont{Itoh}},
  \bibinfo{author}{\bibfnamefont{H.}~\bibnamefont{Sakaguchi}},
  \bibinfo{author}{\bibfnamefont{M.}~\bibnamefont{Uchida}},
  \bibinfo{author}{\bibfnamefont{T.}~\bibnamefont{Ishikawa}},
  \bibinfo{author}{\bibfnamefont{T.}~\bibnamefont{Kawabata}},
  \bibinfo{author}{\bibfnamefont{T.}~\bibnamefont{Murakami}},
  \bibinfo{author}{\bibfnamefont{H.}~\bibnamefont{Takeda}},
  \bibinfo{author}{\bibfnamefont{T.}~\bibnamefont{Taki}},
  \bibinfo{author}{\bibfnamefont{S.}~\bibnamefont{Terashima}},
  \bibinfo{author}{\bibfnamefont{N.}~\bibnamefont{Tsukahara}},
  \bibnamefont{et~al.}, \bibinfo{journal}{Nuclear Physics A}
  \textbf{\bibinfo{volume}{731}}, \bibinfo{pages}{41 } (\bibinfo{year}{2004}),
  ISSN \bibinfo{issn}{0375-9474},
  \urlprefix\url{http://www.sciencedirect.com/science/article/pii/S0375947403018542}.

\bibitem[{\citenamefont{Itoh et~al.}(2003{\natexlab{b}})\citenamefont{Itoh,
  Sakaguchi, Uchida, Ishikawa, Kawabata, Murakami, Takeda, Taki, Terashima,
  Tsukahara et~al.}}]{Itoh2003}
\bibinfo{author}{\bibfnamefont{M.}~\bibnamefont{Itoh}},
  \bibinfo{author}{\bibfnamefont{H.}~\bibnamefont{Sakaguchi}},
  \bibinfo{author}{\bibfnamefont{M.}~\bibnamefont{Uchida}},
  \bibinfo{author}{\bibfnamefont{T.}~\bibnamefont{Ishikawa}},
  \bibinfo{author}{\bibfnamefont{T.}~\bibnamefont{Kawabata}},
  \bibinfo{author}{\bibfnamefont{T.}~\bibnamefont{Murakami}},
  \bibinfo{author}{\bibfnamefont{H.}~\bibnamefont{Takeda}},
  \bibinfo{author}{\bibfnamefont{T.}~\bibnamefont{Taki}},
  \bibinfo{author}{\bibfnamefont{S.}~\bibnamefont{Terashima}},
  \bibinfo{author}{\bibfnamefont{N.}~\bibnamefont{Tsukahara}},
  \bibnamefont{et~al.}, \bibinfo{journal}{Phys. Rev. C}
  \textbf{\bibinfo{volume}{68}}, \bibinfo{pages}{064602}
  (\bibinfo{year}{2003}{\natexlab{b}}),
  \urlprefix\url{https://link.aps.org/doi/10.1103/PhysRevC.68.064602}.

\end{thebibliography}

\end{document}